\documentclass[letter]{aa} 

\usepackage{graphicx}
\usepackage{txfonts}
\usepackage{color}
\usepackage{multirow}
\usepackage{float}
\usepackage{soul}
\begin{document} 

   \title{First detection of H$_2$S in a protoplanetary disk }
   \subtitle{The dense GG Tau A ring}   
    \author {N.T. Phuong \inst{1^\star,2,3}
        \and  E. Chapillon \inst{1,4}
         \and L. Majumdar \inst{5}
         \and A. Dutrey \inst{1}
         \and S. Guilloteau \inst{1}
         \and V. Pi\'etu \inst{4}
         \and V. Wakelam \inst{1}
         \and P. N. Diep \inst{2,3}
         \and Y-W. Tang \inst{6}
         \and T. Beck \inst{7}
        \and J. Bary \inst{8}} 
             
 \institute{Laboratoire d'Astrophysique de Bordeaux, Universit\'e de Bordeaux, CNRS, B18N, All\'ee Geoffroy Saint-Hilaire, F-33615 Pessac; $^\star$thi-phuong.nguyen@u-bordeaux.fr 
                 \and Department of Astrophysics, Vietnam National Space Center, 
                        Vietnam Academy of Science and Techonology, 18 Hoang Quoc Viet, Cau Giay, Hanoi, Vietnam 
                 \and Graduate University of Science and Technology, Vietnam Academy of Science and Techonology, 18 Hoang Quoc Viet, Cau Giay, Hanoi, Vietnam
                 \and IRAM, 300 rue de la piscine, F-38406 Saint Martin d'H\`eres Cedex, France
                 \and Jet Propulsion Laboratory, California Institute of Technology, 4800 Oak Grove Drive, Pasadena, CA 91109, USA
                 \and Academia Sinica Institute of Astronomy and Astrophysics, PO Box 23-141, Taipei 106, Taiwan
                 \and Space Telescope Science Institute, 3700 San Martin Drive, Baltimore, Maryland 21218, USA   
                 \and Department of Physics and Astronomy, Colgate University, 13 Oak Drive, Hamilton, New York 13346, USA}
 \date{Received 04 July 2018; accepted 27 July 2018}

  \abstract
   {Studying molecular species in protoplanetary disks is very useful 
   to characterize the properties of these objects, which are the site 
   of planet formation.} 
   {We attempt to constrain the chemistry of S-bearing molecules
   in the cold parts of circumstellar disk of GG Tau A.}
   {We searched for H$_2$S, CS, SO, and SO$_2$ in the dense disk around 
   GG Tau A with the NOrthem Extended Millimeter Array (NOEMA) interferometer. We analyzed our data using 
   the radiative transfer code DiskFit and the three-phase chemical model 
   Nautilus.} 
   {We detected H$_2$S emission from the dense and cold ring orbiting 
   around GG Tau A. This is the first detection of H$_2$S in a 
   protoplanetary disk. We also detected HCO$^+$, H$^{13}$CO$^+$, and DCO$^+$ in the disk. 
   Upper limits for other molecules, CCS, SO$_2$, SO, HC$_3$N, and $c$-C$_3$H$_2$ are also obtained. The observed DCO$^+$/HCO$^+$ ratio is similar to those in other 
   disks. The observed column densities, derived using our radiative 
   transfer code DiskFit, are then compared with those from our 
   chemical code Nautilus. The column densities are in reasonable 
   agreement for DCO$^{+}$, CS, CCS, and SO$_2$. For H$_2$S 
   and SO, our predicted vertical integrated column densities are more 
   than a factor of 10 higher than the measured values.} 
   {Our results reinforce the hypothesis that only a strong sulfur 
   depletion may explain the low observed H$_2$S column density in 
   the disk. The H$_2$S detection in GG Tau A is most likely linked to the much 
   larger mass of this disk compared to that in other T Tauri systems.}


   \maketitle
%
\section{Introduction}

Understanding the physical and chemical structure of protoplanetary 
disks is needed to determine the initial conditions of planet 
formation. Studies of protoplanetary disks have led to a global 
picture in which disks are flared and layered with important vertical, radial
density, and temperature gradients. The uppermost layer is directly 
illuminated by stellar UV and dominated by photodissociation 
reactions, while molecules stick to dust grains in the very cold midplane. 
In between there is a rich molecular layer \citep{Kenyon+Hartmann_1987, vanZadelhoff+vanDishoeck+Thi+etal_2001}. 
Studies of the gas content rely on trace molecules because H$_2$ is
not detectable at the temperatures of disks.
So far, the molecules that 
have been detected in T Tauri disks are CO, $^{13}$CO, C$^{18}$O, 
C$^{17}$O, CN, CS, H$_2$CO, CCH, DCN, HCO$^+$, H$^{13}$CO$^+$, DCO$^+$, 
N$_2$H$^+$, HC$_3$N, CH$_3$CN, HD, C$_3$H$_2$, C$_2$H$_2$, OH, 
SO, CH$^+$, N$_2$D$^+$, NH$_3$, 
CH$_3$OH, H$^{13}$CN, HC$^{15}$N, C$^{15}$N, and HCOOH \citep{Dutrey+Guilloteau+Guelin_1997, 
Thi+vanDishoeck+Blake+etal_2001, Qi+Wilner+Aikawa_2008, Dutrey+Wakelam+Boehler+etal_2011, 
Chapillon+Dutrey+Guilloteau+etal_2012, Bergin+Cleeves+Gorti_2013, Qi+Oberg+Wilner+etal_2013, 
Huang+Oberg_2015, Oberg+Furuya+Loomis+etal_2015, Walsh+Loomis+Oberg_2016, 
Guilloteau+Reboussin+Dutrey+etal_2016, 
Salinas+Hogerheijde+Bergin+etal_2016, Guzman+Oberg+Loomis_2015, Hily-Blant+Magalhaes+Kastner_2017, Favre+Fedele+Semenov_2018}. 

More than a dozen  S-bearing species have been observed in dense cloud
cores; they are chemically active and often used as chemical clocks in 
low-mass star forming regions \citep{Buckle+Fuller_2003, Wakelam+Caselli+Ceccarelli_2004, Wakelam+Castets+Ceccarelli_2004}. Some S-bearing species, CS, SO, SO$_2$, 
and H$_2$S, are observed in Class 0 and Class I sources 
\citep{Dutrey+Wakelam+Boehler+etal_2011, 
Guilloteau+DiFolco+Dutrey+etal_2013, 
Guilloteau+Reboussin+Dutrey+etal_2016} while CS, the second main 
reservoir of sulfur in the gas phase \citep{Vidal+Loison+Jaziri+etal_2017} 
is the only S-bearing molecule detected in disks around T Tauri stars. 

We report the first detection of H$_2$S in a disk around a T Tauri 
star, GG Tau A. GG Tau, located at 150 pc in Taurus-Auriga  star forming region \citep{GAIA_2016,GAIA_2018}, is a hierarchical quintuple system with the GG Tau A 
triple star \citep[separation $\sim5$ and 
38\,au;][]{DiFolco+Dutrey+LeBouquin_2014} surrounded by a dense ring 
located between 180 and 260\,au and a large disk extending out to 800\,au 
\citep[see][and references therein]{Dutrey+DiFolco+Beck_2016}. The disk 
is massive  (0.15 M$_{\odot}$) and cold; it has a dust temperature of 
14\,K at 200\,au,  a kinetic temperature derived from CO analysis of 
$\sim$20\,K at the same radius \citep{Dutrey+DiFolco+Guilloteau_2014, 
Guilloteau+Dutrey+Simon_1999}, and little or no vertical temperature 
gradient in the molecular layer 
\citep{Tang+Dutrey+Guilloteau+etal_2016}. The large size, low 
temperature, and large mass make GG Tau A disk an ideal laboratory to search for 
cold molecular chemistry. 

Besides the H$_2$S detection, we also report detections of HCO$^+$, DCO$^+$, and 
H$^{13}$CO$^+$ and discuss the upper limits of CCS, 
SO$_2$, SO, c-C$_3$H$_2$, and HC$_3$N. 
\section{Observations and results} 
\subsection{Observations} 

The H$_2$S 1(1,0)-1(0,1) observations were 
carried out with the NOrthem Extended Millimeter Array (NOEMA) on 23 
December, 2017 using D configuration with nine antennas. The total on 
source integration time is 5.2 hours. Baselines ranging between 24\,m 
and 176\,m provide an angular resolution of $2.50"\times1.9"$, 
PA=15$^\circ$. Phase and amplitude calibrations were performed using 
0507+179 and 0446+112. Flux calibration was carried out using MWC349 as a 
reference (flux 1.6 Jy at 170.3 GHz). 
The full 7.74 GHz upper and lower sidebands of the new PolyFiX correlator were covered at 2 MHz channel spacing, and high spectral resolution (62.5 kHz) windows covered 
lines of H$_2$S 1(1,0)-1(0,1), H$^{13}$CO$^+$ (2-1), CCS, SO$_2$, SO, HC$_3$N, and c-C$_3$H$_2$.

DCO$^+$ (3-2) was observed with PdBI interferometer (now known as NOEMA) in December 2013 and April 2014 with six antennas
at an angular resolution of $1.76"\times1.23"$, PA=17$^\circ$.
Phase and amplitude calibrations were performed using 0507+179 and 
0446+112, while the flux calibration was carried out using 3C84 and MWC\,349. 

The HCO$^+$(1-0) data are from \citet{Guilloteau+Dutrey+Simon_1999} and are
processed in this work with a resolution of $4.57"\times2.55"$, at PA=$-38^\circ$. We used the GILDAS{\footnote{\tt https://www.iram.fr/IRAMFR/GILDAS/}}  software package 
to reduce the data. Images were 
produced using natural weighting and {\it{Hogbom}} algorithm. The 
continuum emission is subtracted from the line maps. 

\begin{figure*}[htbp!]
   \centering  
\mbox{ \includegraphics[width=4.1cm, trim=2cm 2cm 8cm 2cm]{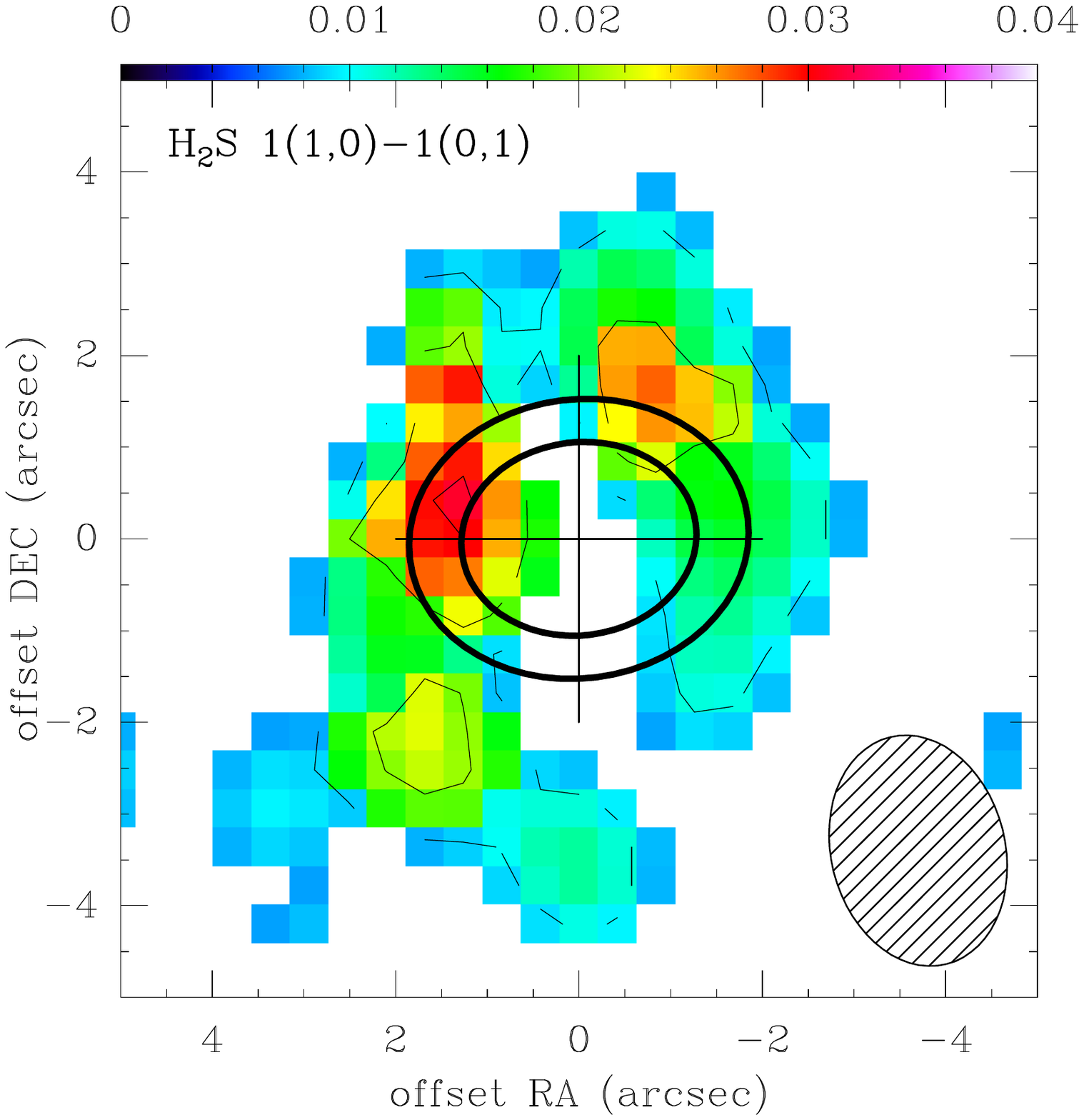}
  \includegraphics[width=4.1cm, trim=2cm 2cm 8cm 2cm]{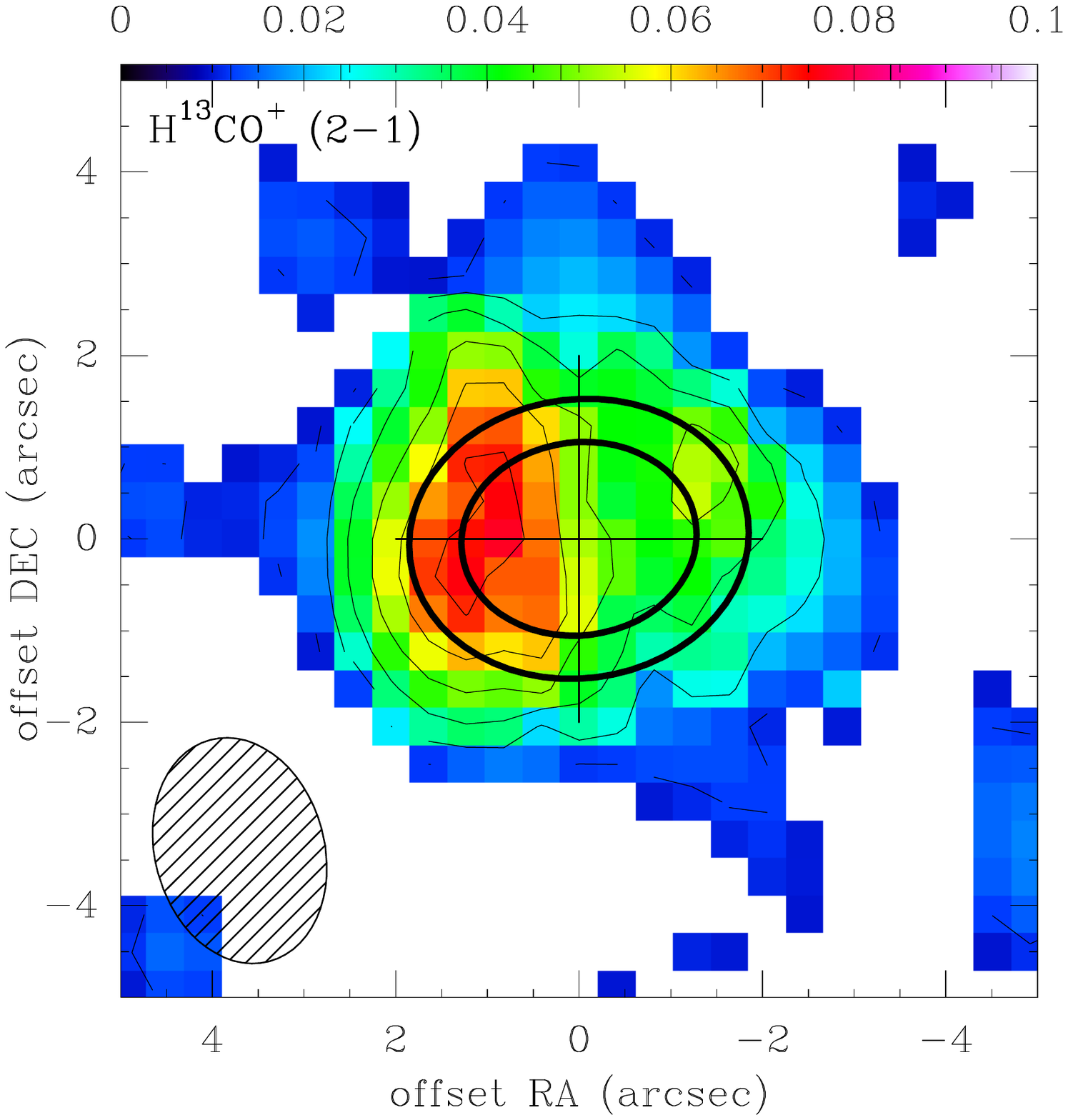} 
  \includegraphics[width=4.1cm, trim=2cm 2cm 8cm 2cm]{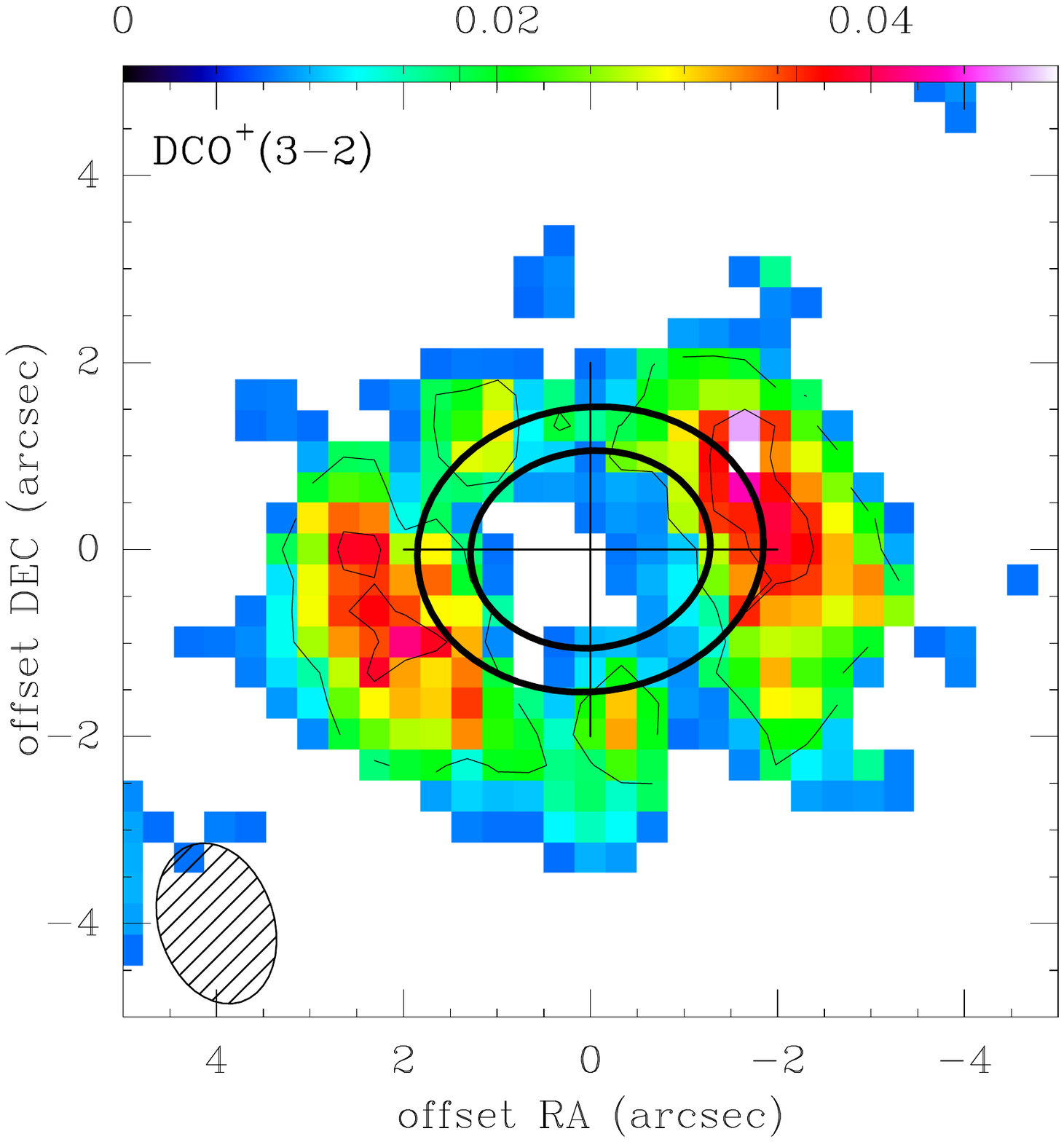}   
  \includegraphics[width=5.cm, trim=2cm 2cm 8cm 2cm]{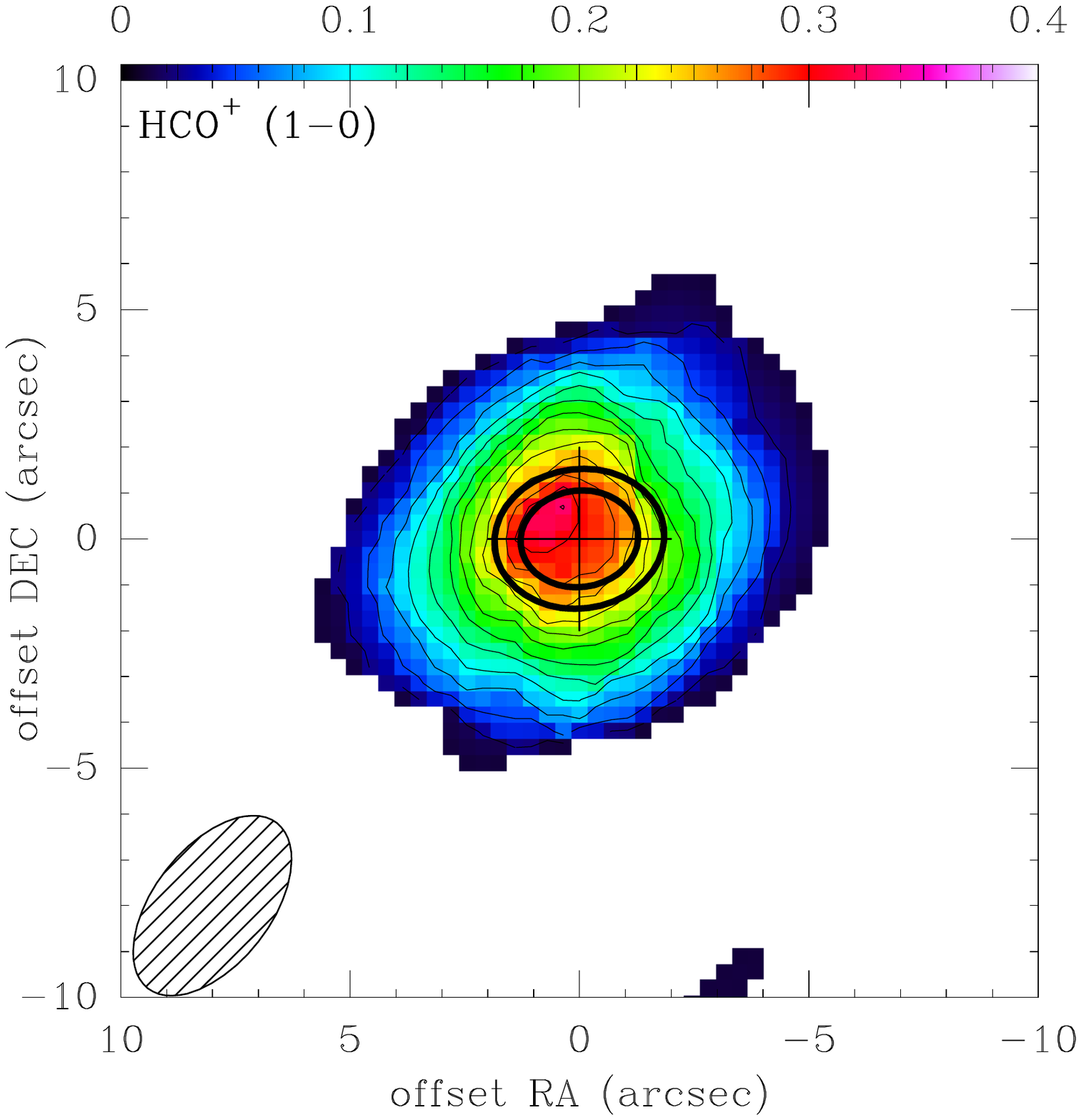}}
\mbox{ \includegraphics[width=4.1cm, trim=2cm 2cm 8cm 2cm]{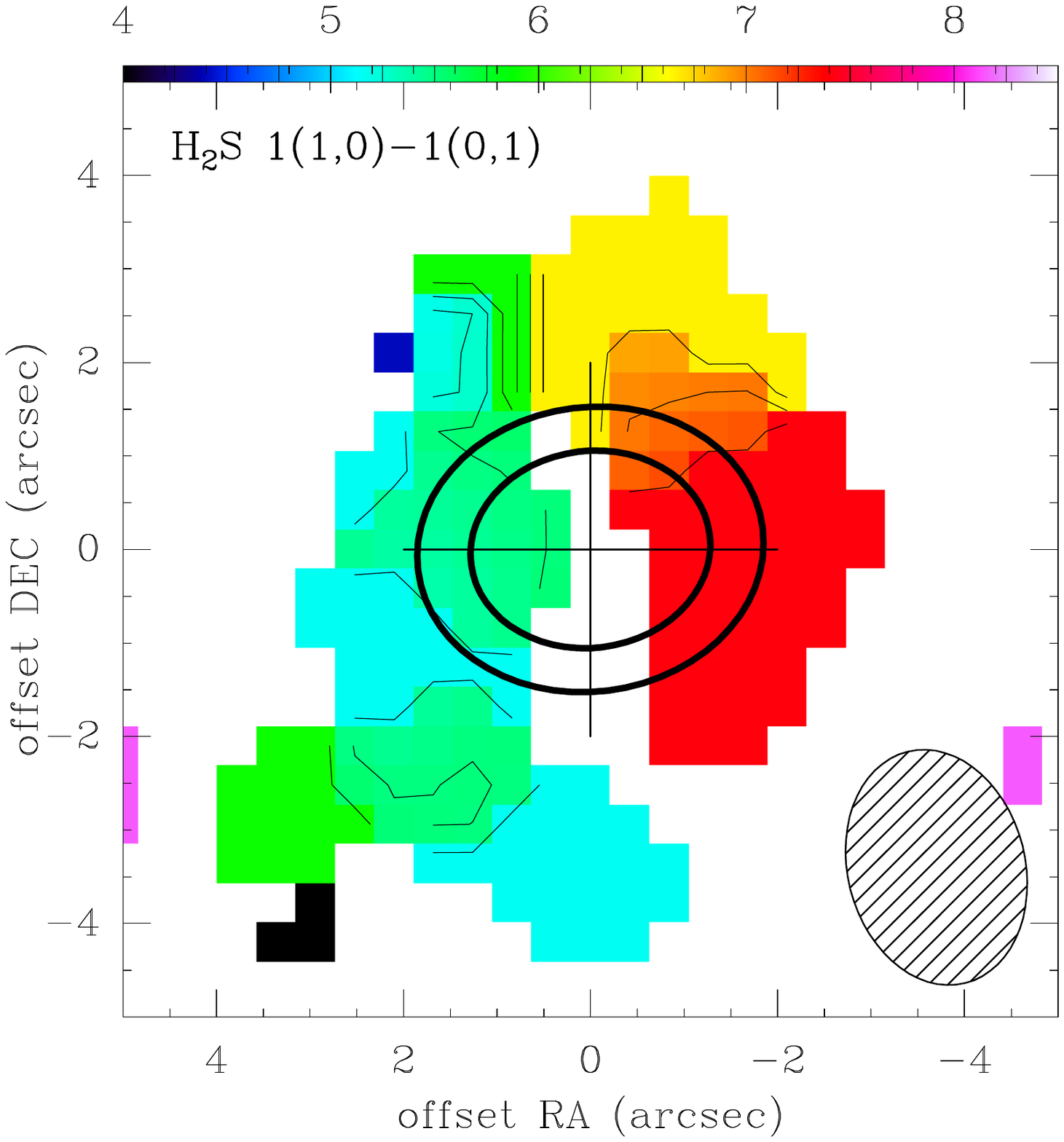}
   \includegraphics[width=4.1cm,trim=2cm 2cm 8cm 2cm]{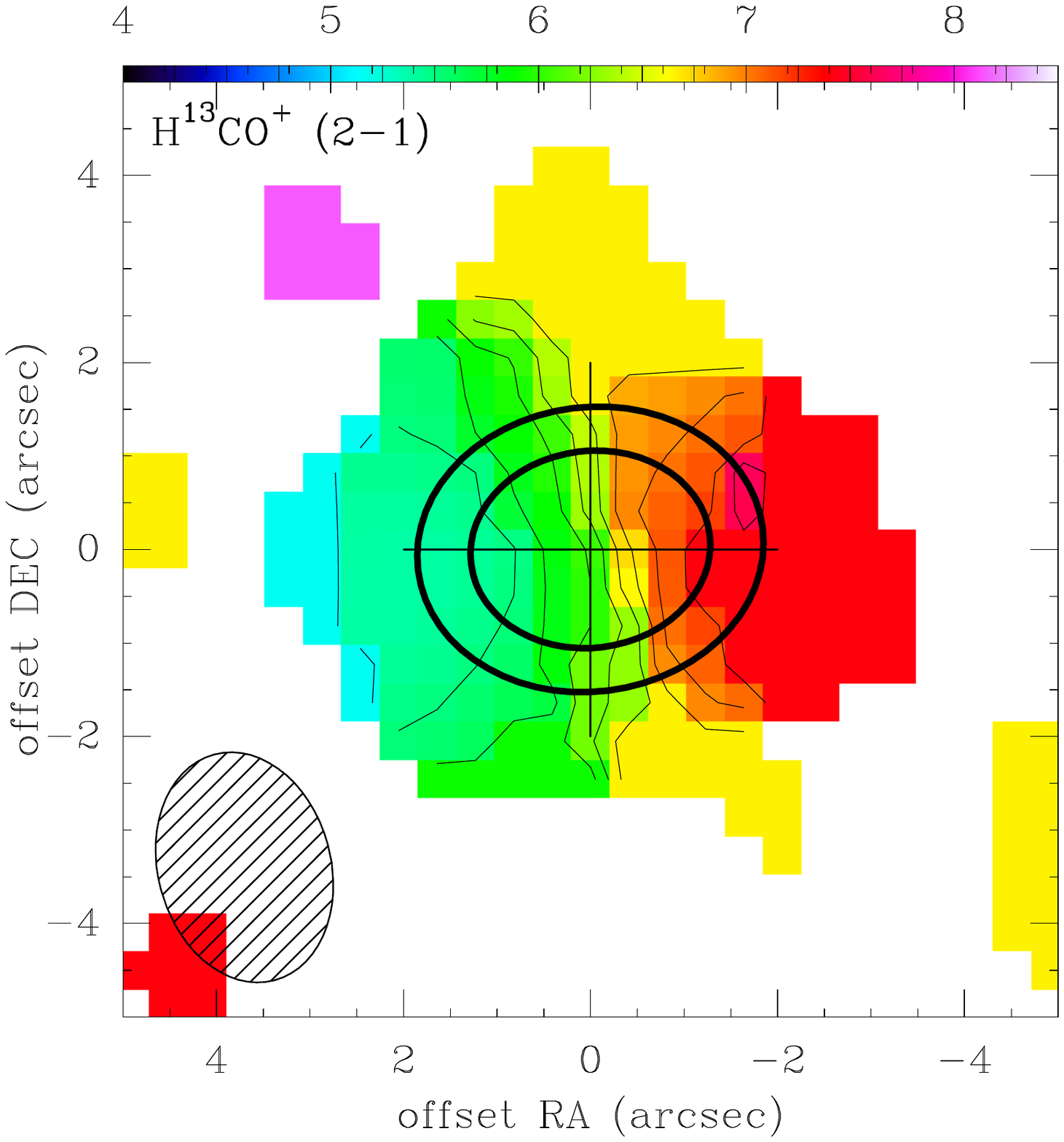}
   \includegraphics[width=4.1cm,trim=2cm 2cm 8cm 2cm]{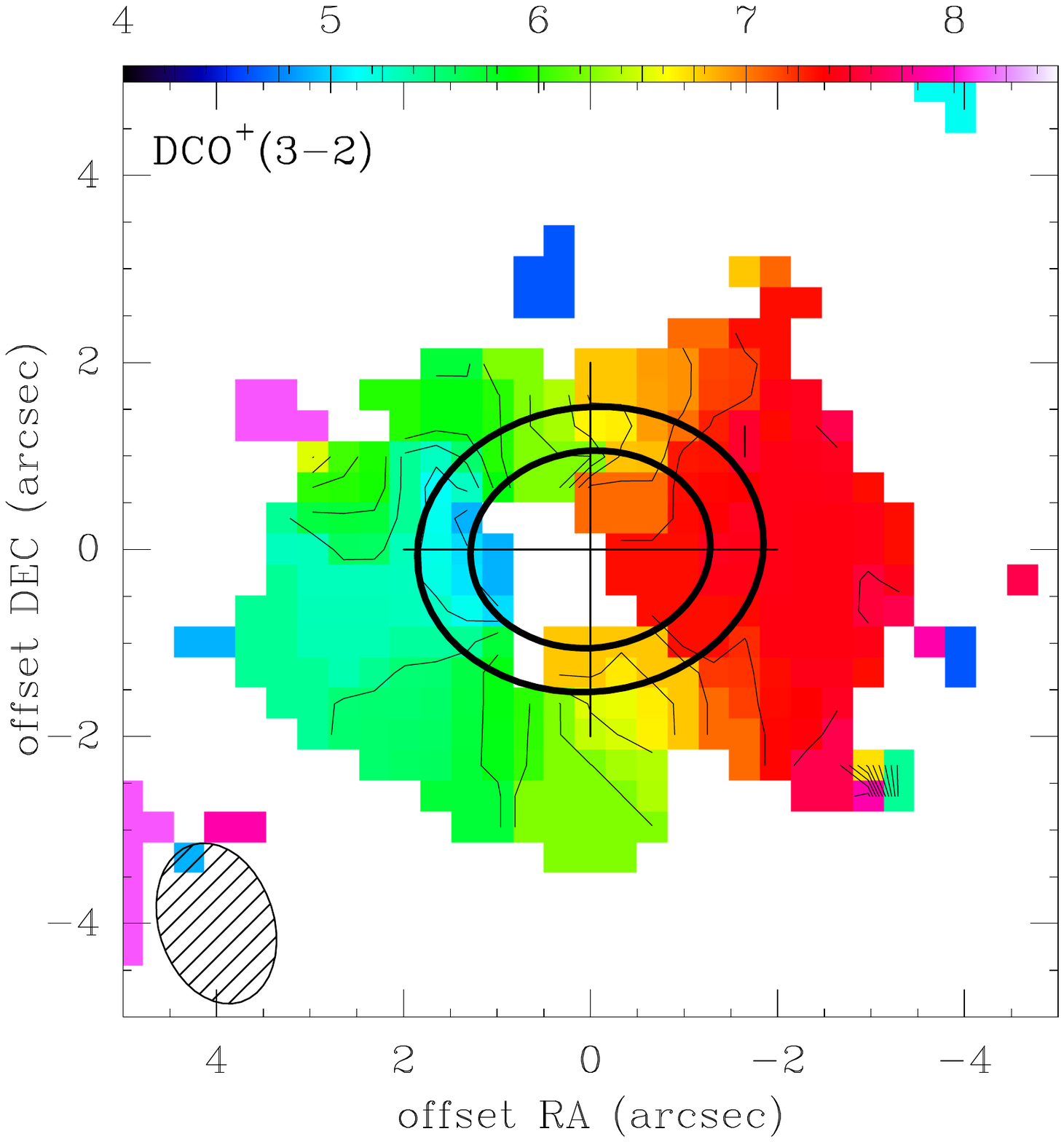}  
   \includegraphics[width=5.1cm,trim=2cm 2cm 8cm 2cm]{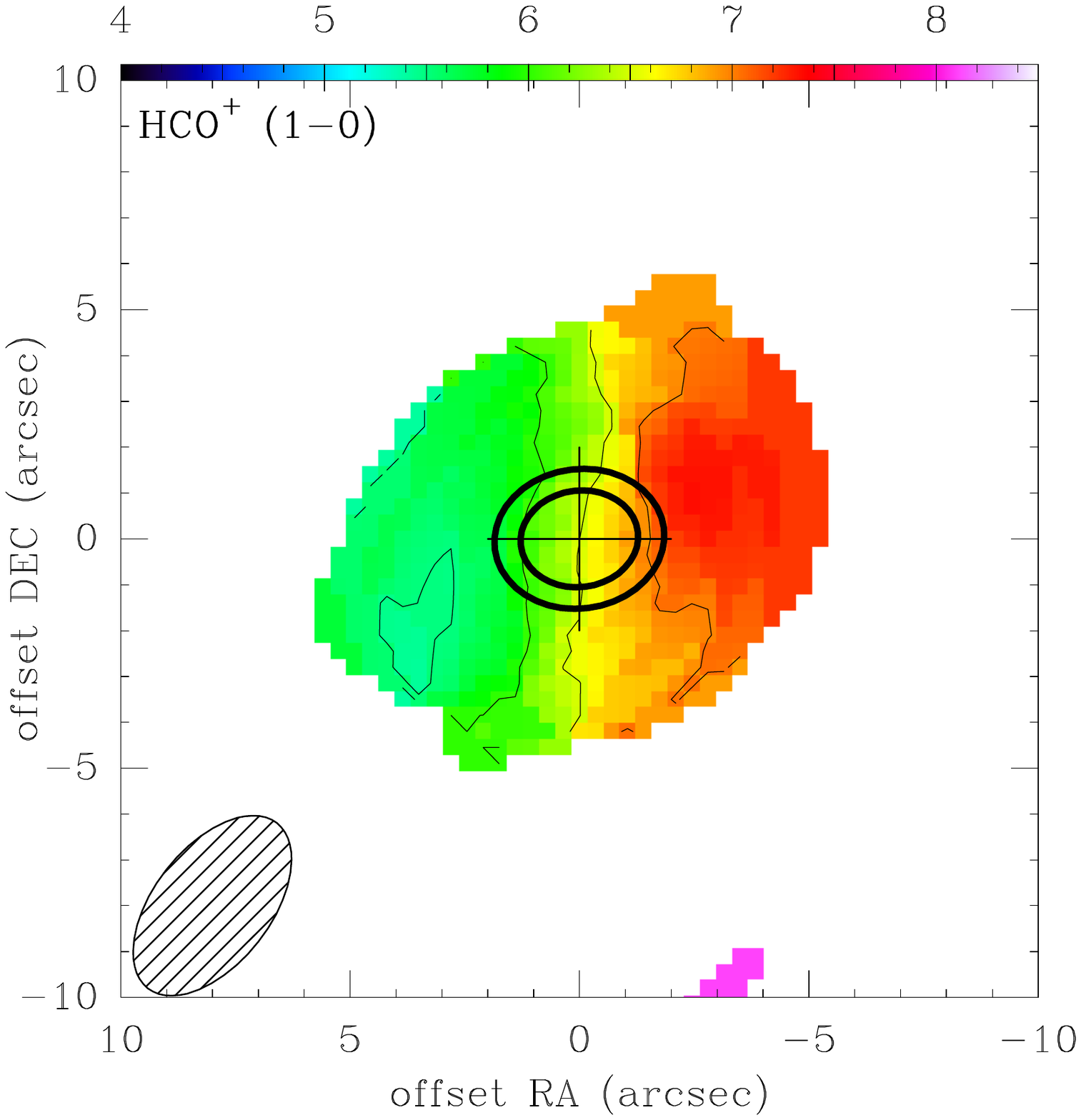}}
%
   \caption{{\it{Upper:}} Integrated intensity maps. The  color scale  
   is in the unit of (Jy\,beam$^{-1}$\,km\,s$^{-1}$). Contour level step is $2\sigma$. {\it{Lower:}} Velocity maps.
   Contour level step is 0.5 km\,s$^{-1}$. Beam sizes are indicated. The ellipses show
   the location of inner ($\sim$180\,au) and outer ($\sim$260\,au) radii of the dust ring.} \label{Fig1}
\end{figure*}
\subsection{Results}
Figure \ref{Fig1} shows integrated intensity maps (upper panels) and velocity 
maps (lower panels) of the detected lines, \mbox{H$_2$S 1(1,0) - 1(0,1)}, 
H$^{13}$CO$^+$ (2-1), DCO$^+$(3-2), and HCO$^+$(1-0). 
The velocity maps show a clear signature of Keplerian rotation.
Channel maps are presented in Appendix \ref{apen:chan}.

H$_2$S\,$1(1,0)-1(0,1)$ is clearly detected with a peak \mbox{S/N$\gtrsim$4 in several channels}.
Most of the line emission originates 
from the dense ring between 180 to 260\,au and extends up 
to $\lesssim$500\,au. The weak east-west asymmetry is unlikely to
be significant given the limited signal-to-noise ratio (S/N).

HCO$^+$(1-0) and H$^{13}$CO$^+$(2-1) are detected with high 
S/N\mbox{($\ge$ 7)}. HCO$^+$(1-0) is as extended as the CO emission out to $\sim$800\,au 
\citep{Guilloteau+Dutrey+Simon_1999}. The optically thin emission from the J=2-1 line of the H$^{13}$CO$^+$ 
isotopolog peaks on the dense ring and extends to $\sim$500\,au. 
On the contrary, the DCO$^+$(3-2) emission, detected at \mbox{S/N$\ge$ 7}, peaks just outside of
the dense ring, near 280 au, suggesting radially varying deuteration.
Other sulfur-bearing species, SO, SO$_2$, CCS, and carbon-bearing 
species HC$_3$N and c-C$_3$H$_2$, are not detected.
\section{Data analysis}

\subsection{DiskFit modeling} 
We assume the physical parameters that govern line emission to vary as 
power laws of the radii 
\citep{Dutrey+Guilloteau+Simon_1994, 
Pietu+Dutrey+Guilloteau+etal_2007}. The data were analyzed inside the 
$uv$ plane using the radiative transfer code DiskFit, which uses $\chi^2$ 
minimization technique, comparing the observed visibilities to 
visibilities predicted by ray tracing  
\citep{Pietu+Dutrey+Guilloteau+etal_2007}.  
%
\begin{table}[H] 
\caption{GG Tau parameters}             
\label{table:disk}      
\centering                          
\begin{tabular}{|c  c| l |}        
\hline              
\multicolumn{2}{|c|}{Geometry} & \multicolumn{1}{c|}{Law}\\ 
\hline 
Inclination & 35$^\circ$ & $V(r)=3.4\,(\frac{r}{100\,\textnormal{au}})^{-0.5}$ \, (km\,s$^{-1}$) \\ 
Orientation & 7$^\circ$ & $T(r)=25\,(\frac{r}{200\,\textnormal{au}})^{-1}$\,\,\,\,\,\,\,\,\,\,\,\, (K)\\ 
Systemic velocity & 6.4 km s$^{-1}$ & $\Sigma(r)=\Sigma_{250}\,(\frac{r}{250\,\textnormal{au}})^{-1.5}$\, (cm$^{-2}$)  \\ 
\hline 
\end{tabular}
\end{table}

The source parameters such as geometry (inclination, orientation, 
and systemic velocity), velocity, and temperature power laws are kept constant 
as they are well known from previous studies \citep[][and our \mbox{Table 
\ref{table:disk}}]{Dutrey+Guilloteau+Simon_1994, 
Dutrey+DiFolco+Guilloteau_2014, Guilloteau+Dutrey+Simon_1999}. Only the molecule surface density parameter $\Sigma_{250}$ was varied during the 
minimization process. 
Results are presented in Table \ref{table:dens} with $3 \sigma$ upper
limits for undetected molecules. 
\subsection{Nautilus modeling}
To model the chemistry in the dense and cold ring of GG Tau A, we used the gas-grain chemical model {\tt Nautilus} 
\citep{2016MNRAS.459.3756R}. This model simulates chemistry in three phases, 
i.e., gas phase, grain surface, and grain mantle, along with possible 
exchanges between the different phases. The reference chemical network 
is {\tt deuspin.kida.uva.2016}  \citep{2017MNRAS.466.4470M} with the 
updates in sulfur chemistry from \cite{Vidal+Loison+Jaziri+etal_2017}. 
The disk 
structure is similar to that used in \cite{Wakelam+Ruaud+Hersant_2016}. 
In addition to disk parameters from Table \ref{table:disk}, we assume a stellar UV 
flux of $\rm f_{UV200AU}$ = 375 $\chi_0$ at 200\,au, where $\chi_0$ is in 
the units of the \cite{1978ApJS...36..595D} interstellar UV field, based on what is observed in T Tauri stars \citep{Bergin+Calvet+Sitko_2004}. 
Based on the observation \citep{Tang+Dutrey+Guilloteau+etal_2016}, we introduced a small vertical temperature gradient with T$_k$=30\,K at three scale heights.
\begin{table*}
\caption{Observed and predicted surface densities (cm$^{-2}$)}             
\label{table:dens}      
\centering                          
\begin{tabular}{|c c c || c c c|}          
\hline   
\multicolumn{3}{|c||}{Detection} & \multicolumn{3}{c|}{Non-detection} \\
\hline
Molecule &  Observed$^\star$  & Predicted$\dagger$ &  Molecule & Observed$^\star$ &  Predicted$\dagger$ \\
               &  (derived from DiskFit)               & (from Nautilus)  & &(derived from DiskFit) &  (from Nautilus)\\ 
\hline 
  HCO$^+$(1-0)  & 1.5 $\pm$ 0.04 $\times$10$^{13}$ & 2.2 $\times$10$^{12}$    & CCS&$<$1.7 $\times$ 10$^{12}$ & 7.2 $\times$10$^{10}$  \\
  H$^{13}$CO$^+$ (2-1) & 5.3  $\pm$ 0.3$ \times$10$^{11}$ & (-)& SO$_2$ & $<$1.5 $\times$10$^{12}$ & 6.0 $\times$10$^{12}$\\ 
  DCO$^+$ (3-2) & 3.9 $\pm$ 0.2 $\times$10$^{11}$ & 7.0 $\times$10$^{10}$        & SO &$<$1.1 $\times$ 10$^{12}$ & 1.5 $\times$10$^{13}$ \\ 
  H$_2$S 1(1,0) - 1(0,1) & 1.3 $\pm$ 0.1 $\times$10$^{12}$ &  3.4 $\times$10$^{13}$        & HC$_3$N&$<$ 3.2 $\times$10$^{11}$ & 5.7 $\times$10$^{11}$ \\
  CS(3-2) &  2.2 $\times$ 10$^{13}$$^\#$ &  1.4 $\times$10$^{13}$     & c-C$_3$H$_2$&$<$ 2.7 $\times$10$^{11}$ & 2.4 $\times$10$^{12}$ \\
 \hline                                  
\end{tabular}
\tablefoot{$^\star$ Observed surface density at 250\,au is derived using DiskFit.
$^\#$ Phuong et al., in prep.\\ 
{$^\dagger$ Species surface density in the gas phase at 250\,au predicted with Nautilus.
$(-)$ Our model does not include carbon isotope chemistry.}} 
\end{table*}
%
\begin{figure*}[htbp!]
   \centering
  \includegraphics[width=3.cm, angle=90]{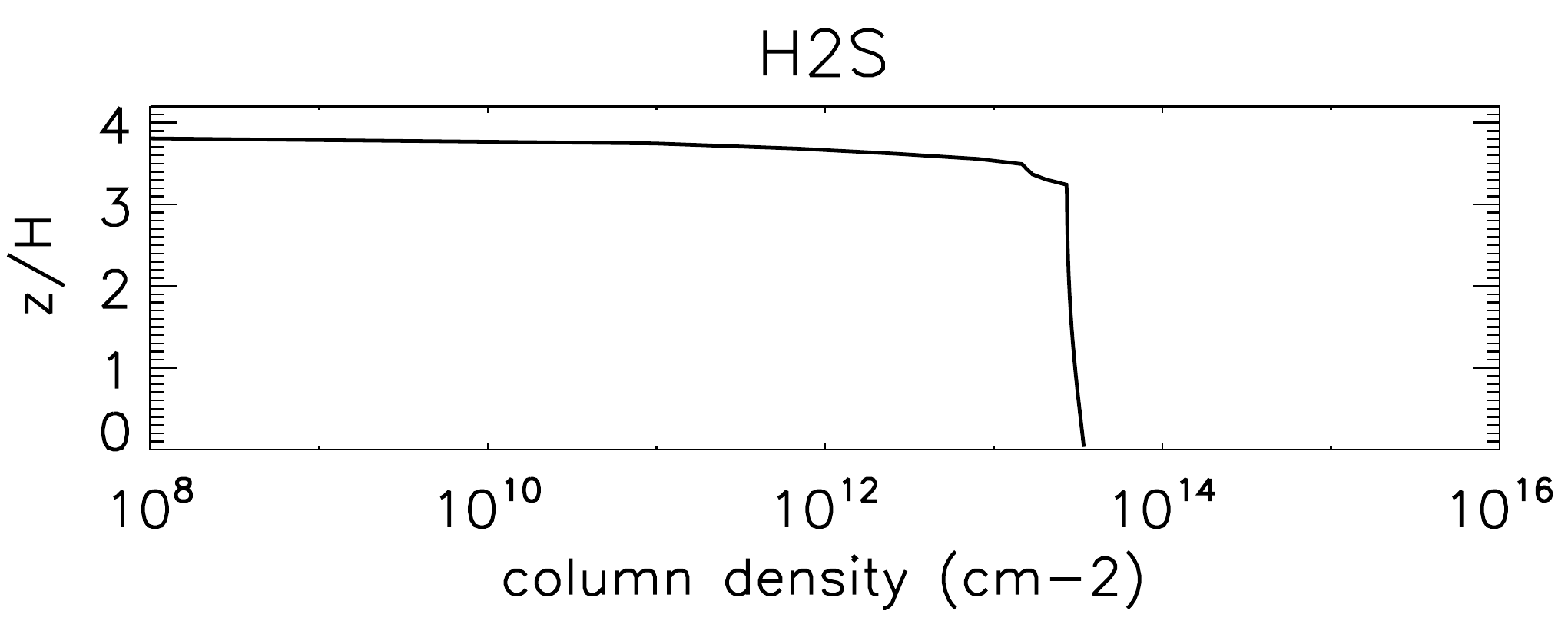}
  \includegraphics[width=3.cm, angle=90]{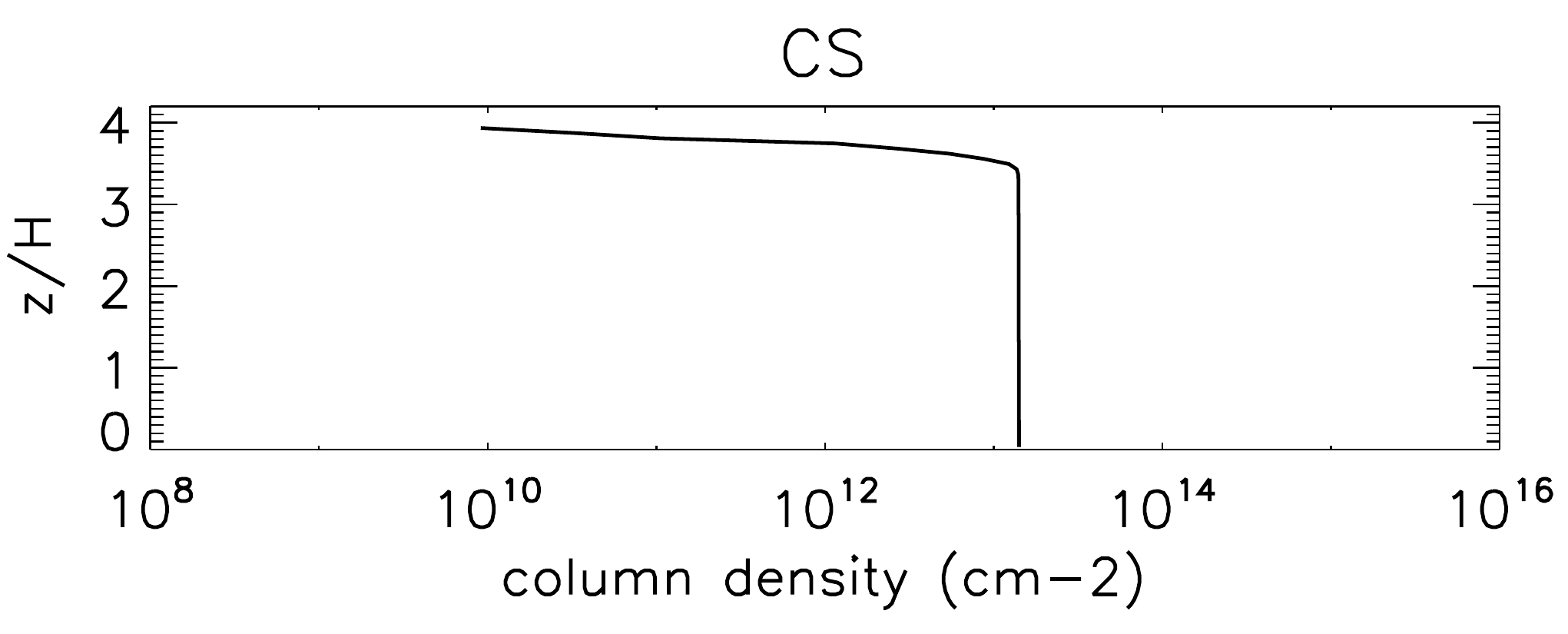}
  \includegraphics[width=3.cm, angle=90]{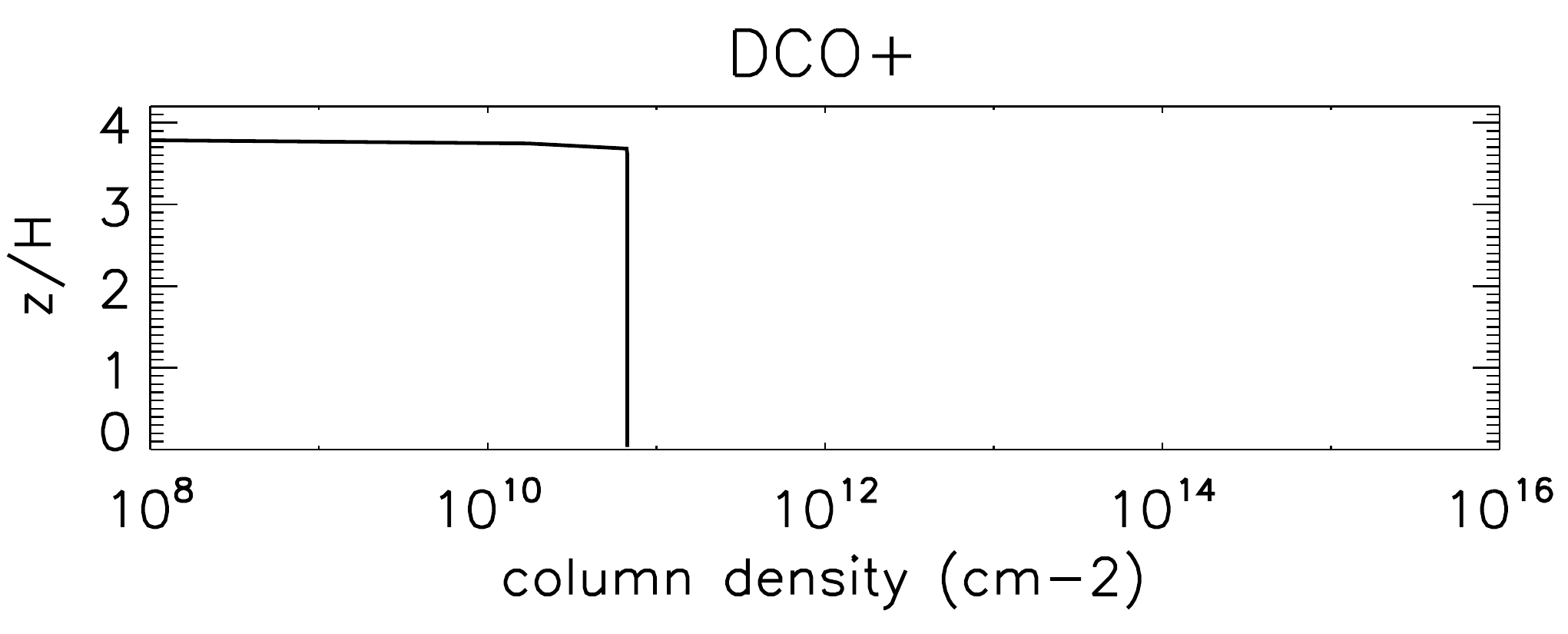}
  \includegraphics[width=3.cm, angle=90]{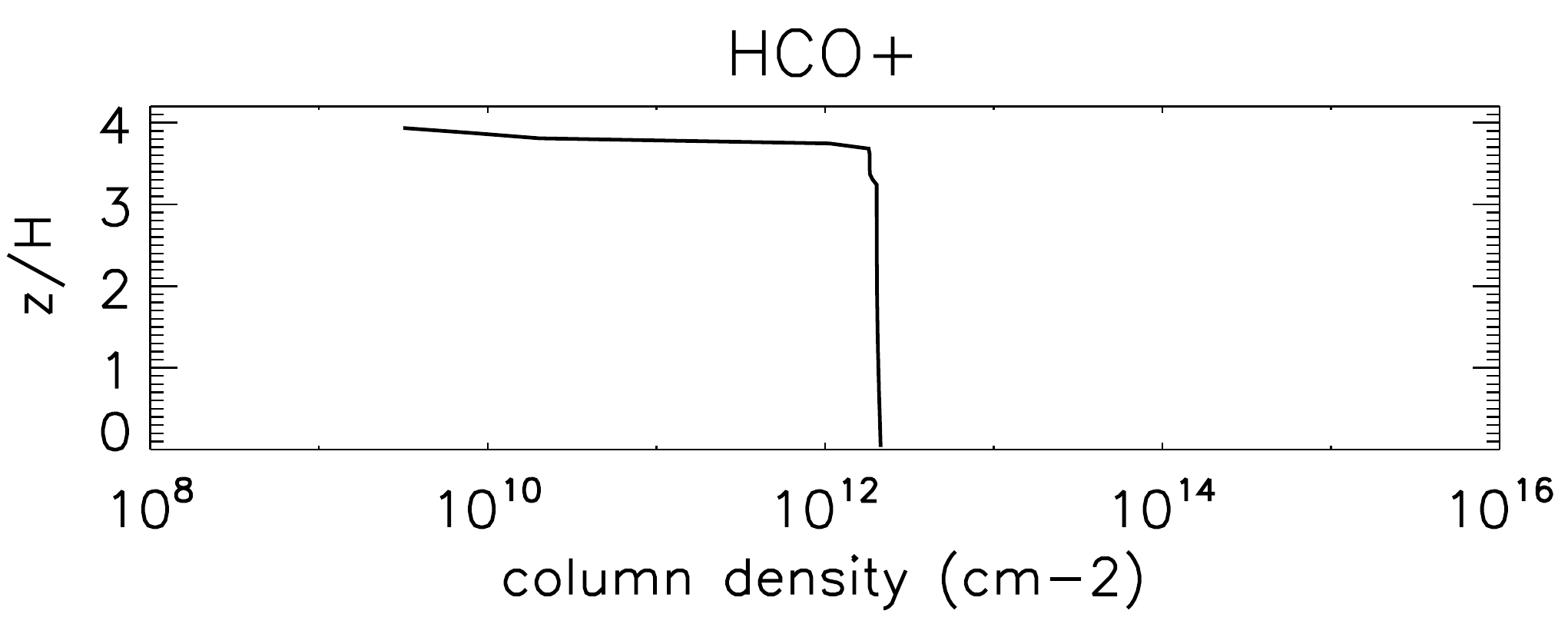}
   \caption{Best model of H$_2$S, CS, DCO$^+$, and HCO$^+$ in the GG 
   Tau A ring derived from Nautilus.  The surface density is shown vs. the z/H ratio (z/H=1 means 1 scale height). } \label{fig:model}
\end{figure*}    

To compute the chemistry, we first calculated the chemical composition 
of the gas and ices of the parent cloud, assuming conditions for a 
dense cloud with an age of $\sim$10$^6$ yr and then ran the model for another 10$^6$ yr 
\citep{Wakelam+Ruaud+Hersant_2016}. For the parent cloud, initially all 
the elements are in atomic form \citep[see Table 1,][]{Vidal+Loison+Jaziri+etal_2017} except for hydrogen and deuterium, 
which are initially in H$_2$ and HD forms, respectively 
\citep{2017MNRAS.466.4470M}. 

We present the trends of the chemistry inside the ring at a radius 
of 250 au in order to explain the observed column densities of H$_2$S, 
CS, DCO$^+$, and HCO$^+$. We explored various initial C/O ratios, ortho to para ratios 
for H$_2$ (OPR), initial sulfur abundances X(S), grain sizes, and UV 
flux. According to \citet{Bergin+Du+Cleeves_2016}, CCH emission can only be explained with a gas-phase C/O ratio larger than 1. This represents a scenario in which oxygen is depleted on the grains before the formation of the disk and driven to the midplane of the disk.  In other words, oxygen would not participate in the chemistry in the region where they observe CCH. \citet{Semenov+Favre+Fedele_2018} found 
that the column densities of SO and SO$_2$ drop by factors of $\sim$100 and 500, respectively, when C/O changes from 0.46 to 1.2, whereas column densities of H$_2$S do not change as the species 
contains neither C nor O. We stick to the standard C/O ratio of 0.7 in our model \citep{Hincelin+Wakelam+Hersant_2011, Wakelam+Ruaud+Hersant_2016, 2017MNRAS.466.4470M}, which
gives a reasonably good agreement for DCO$^+$, CS, CCS, HC$_3$N, and SO$_2$. 

Results are therefore presented for C/O\,=\,0.7, OPR=3, X(S)=$8\,10^{-8}$ 
and a grain size of 0.1 $\mu$m. Other models lead to larger disagreement
with the data. Figure \ref{fig:model} and \ref{apen:chem} show the 
predicted vertical distribution of the molecules, and Table 
\ref{table:dens} compares the predicted surface densities to the 
observational results derived using DiskFit.
%
\section{Discussion}
\subsection{Comparison with other sources} 
The measured H$_2$S column density is a factor of three greater than
the upper limits quoted by \citet{Dutrey+Wakelam+Boehler+etal_2011} 
for DM\,Tau, LkCa\,15, MWC\,480, and GO\,Tau, probably reflecting
the larger disk mass of GG Tau A. However, the CS to H$_2$S abundance ratio 
of $\sim$20 in GG Tau A may still be similar in all sources.
The upper limit on HC$_3$N is about two times
lower than the detections reported in LkCa\,15, MWC\,480, and GO\,Tau by
\citet{Chapillon+Dutrey+Guilloteau+etal_2012}.

To make relevant abundances comparisons, we use $^{13}$CO as a reference since H$_2$ column densities are difficult to 
accurately determine. The results for the disks of GG Tau A and LkCa15 and 
the dark cloud TMC-1 are given in Table \ref{table:abun}. LkCa15 is a T Tauri star similar to GG Tau A: its disk exhibits a central cavity of 
radius 50 au \citep{Pietu+Dutrey+Guilloteau_2006} and has a mass on the order of $\sim 0.028$ M$_{\odot}$ \citep{Guilloteau+Dutrey+Pietu_2011}. Determining the uncertainties is difficult because the abundances were obtained from different studies. Therefore, we assume errors of 30\% in the 
cases of LkCa 15 and TMC-1.

For GG Tau A, we take a $^{13}$CO column density, derived from 
observations, at 250\,au of $\Sigma_{250}$=1.13 10$^{16}$ cm$^{-2}$ 
(Phuong et al., in prep). 
For LkCa 15, 
\citet{Punzi+Hily-Blant+Kastner+etal_2015} found HCO$^+$ abundance 
relative to $^{13}$CO of 15\,10$^{-4}$, 
\citet{Huang+Oberg+Qi+etal_2017} gave abundance ratios of 
DCO$^+$/HCO$^+$ and DCO$^+$/H$^{13}$CO$^+$ of 0.024 and 1.1, 
respectively, and \citet{Dutrey+Wakelam+Boehler+etal_2011}  
gave an upper limit of H$_2$S relative to CO of 10$^{-6}$,
which we convert to $^{13}$CO using an isotopic ratio
$^{12}$C/$^{13}$C\,$\sim60$ \citep{Lucas+Liszt_1998}.

In the TMC-1 dark cloud, \citet{Ohishi+Irvine+Kaifu_1992} determined 
$^{12}$CO abundance relative to H$_2$ of 8\,10$^{-5}$ or 1.3\,10$^{-6}$ for $^{13}$CO. 
The abundance  relative to H$_2$ of HCO$^+$, H$_2$S (upper limit) \citep{Omont_2007}, 
H$^{13}$CO$^+$, and DCO$^+$ \citep{Butner+Lada+Loren_1995} are then used 
to get the abundances relative to $^{13}$CO. 
In L134N, the abundances of these species are similar, but H$_2$S has been detected with an
abundance ratio of 60\,10$^{-5}$ \citep{Ohishi+Irvine+Kaifu_1992}, similar to the upper limit obtained
in TMC-1.
Thus, the disks appear to have very similar relative abundances, suggesting
similar chemical processes at play, while the dense core differs significantly.

 \begin{table}
\caption{Molecular abundance relative to $^{13}$CO (X$_{[mol]}$/X$_{[^{13}CO]}\times10^5$)}         
\label{table:abun}      
\centering                          
\begin{tabular}{|c|c|c|c|c|}        
\hline              
& TMC-1$^\star$  & LkCa 15 & GG Tau \\
\hline
HCO$^+$ & $600\pm180$$^{(1)}$ & $150\pm35$$^{(3)}$ &$130\pm12$ \\
\hline
H$_2$S& $<45^{(1)}$& $<7^{(4)}$ &$11\pm3$\\
\hline   
H$^{13}$CO$^+$& $15\pm4$ $^{(2)}$ & $5\pm1.5$ $^{(4)}$ & $4.7\pm0.3$\\
\hline
DCO$^+$ & $30\pm9$ $^{(2)}$ & $4.5\pm1.4$ $^{(4)}$ & $3.5\pm0.15$  \\
\hline                             
\end{tabular}
\tablefoot{$^\star$ $^{13}$CO abundance is derived from CO abundance in 
\citet{Ohishi+Irvine+Kaifu_1992}, $^{(1)}$ \citet{Omont_2007}, $^{(2)}$ 
\citet{Butner+Lada+Loren_1995}, $^{(3)}$ 
\citet{Punzi+Hily-Blant+Kastner+etal_2015}, $^{(4)}$ 
\citet{Dutrey+Wakelam+Boehler+etal_2011}, $^{(5)}$ 
\citet{Huang+Oberg+Qi+etal_2017}.}
\end{table}

\subsection{Sulfur-bearing species}

In the chemical modeling, we found that H$_2$S peaks around three scale heights. 
The main reason behind this is 
rapid formation of H$_2$S on the grain surface  via the hydrogenation 
reaction of HS, i.e., grain-H + 
grain-HS$\rightarrow$grain-H$_2$S. Once H$_2$S is formed on the 
surface, it is then chemically
desorbed to the gas phase. Almost 80\% of the H$_2$S comes from the 
surface reactions, whereas the contribution of the gas-phase reaction 
H$_3$S$^+$+e$^-$$\rightarrow$ H + H$_2$S is about 20\%. Below three scale heights, 
H$_2$S depletes rapidly on the grains because of
the increase in density and decrease in temperature. At the same altitude,
CS is formed in the gas phase via the dissociative 
recombination reactions of HCS$^+$, H$_2$CS$^+$,  H$_3$CS$^+$, and 
HOCS$^+$. 

The modeled  CCS and SO$_2$ column densities (shown in Table \ref{table:dens} and in 
Appendix \ref{apen:chem}) are low, explaining their non-detection but the SO column density is overpredicted. 
The CCS molecule shows its peak above $z/H$=3 and is due to the gas phase formation 
via \mbox{S + CCH$\rightarrow$ H + CCS} and \mbox{HC$_2$S$^+$ + e$^-$ $\rightarrow$ 
H + CCS} reactions. SO$_2$ is made from the OH + SO reaction around this 
location, whereas SO comes from the S + OH reaction.

We found that the UV field has a negligible impact on the H$_2$S desorption
and mildly affects the SO/H$_2$S ratio. The key parameter in the
model is the initial S abundance. Even with the low value, $8\,10^{-8}$,
the chemical model overpredicts H$_2$S and SO by about an order
of magnitude, but is compatible with CS and the current limits on SO$_2$ and CCS. 

In our models, the molecular layer is very thin and situated high
above the disk plane at three scale heights. This is at odds with the 
observations of CS in the Flying Saucer \citep{Dutrey+Guilloteau+Pietu_2017}, where
CS appears closer to one scale height. The difference may be due to the larger mass
of the GG Tau disk (0.15 M$_{\odot}$). On one side, the high densities limit the UV 
radiation penetration (which drives the active chemistry) to the
uppermost layers, while closer to the midplane, the even higher densities
lead to more efficient depletion on dust grains.

Our results suggest that chemistry for H$_2$S 
on the grain surfaces is likely not properly taken into account even with 
our three-phase model and that a significant amount of H$_2$S should 
change in some more complex sulfur-bearing species, limiting the
overall desorption of S-bearing molecules  
\citep{Dutrey+Wakelam+Boehler+etal_2011, 
Wakelam+Ceccarelli+Castets_2005}. Indeed, measurements of S-bearing species in comets 67P performed by ROSETTA indicate a solar value
for the S/O elemental ratio within 2$\sigma$ errors \citep{Calmonte+Altwegg+Balsiger_2016}. 
H$_2$S accounts for about half of the S budget in the comet, suggesting that transformation of H$_2$S
to other compounds in ices is limited. The nearly constant H$_2$S/H$_2$O ratio also suggests that H$_2$S
does not evaporate alone, but in combination with water \citep{Jim+Caro_2011}.

\subsection{Chemistry of DCO$^+$ and other observed species}

\paragraph {Chemistry of DCO$^+$:} The measured HCO$^+$/H$^{13}$CO$^+$ ratio is about 30, suggesting partially 
optically thick emission for HCO$^+(1-0)$ line. The measured 
DCO$^+$/HCO$^+$ ratio, $\sim0.03$ over the disk, is comparable to the 
averaged value \citep[$\sim$0.04;][]{vanDishoeck+Thi+vanZadeldoff+etal_2003} derived in the  disk of TW Hydra 
of mass of $\sim$ 0.06 M$_{\odot}$ \citep{Bergin+Cleeves+Gorti_2013}, and in the disk of 
LkCa\,15 \citep[ratio of $\sim$0.024,][]{Huang+Oberg+Qi+etal_2017}. This shows 
clear evidence of ongoing deuterium enrichment.  

HCO$^+$ formation and deuteration is controlled by CO as well as
 H$_2$D$^+$ and H$_3^+$ ions. These ions are mostly sensitive
to the X-ray flux, while UV radiation and cosmic rays play a limited
role, and their balance is controlled by the temperature sensitive
reaction \mbox{H$_3^+$+HD $\rightleftharpoons$ H$_2$D$^+$+H$_2$}.  
Because of the temperature dependences, DCO$^+$ is expected to
be enhanced around the CO snow-line, as illustrated by the ring
structure in HD 163296 \citep{Mathews+Klaasen+Juhasz+etal_2013}.
Our model somewhat underpredicts the HCO$^+$ content.  At 250 au, 
HCO$^+$ peaks at three scale heights, where the molecular layer is warm 
($\sim$ 30\,K) and forms mainly from the reaction of \mbox{CO + 
ortho-H$_3$$^+$}. At this altitude, DCO$^+$ forms from the isotope 
exchange reaction between HCO$^+$ and D because the gas temperature is 
still high. Closer to the disk midplane, the \mbox{ortho-H$_2$D$^+$ + CO} pathway 
remains inefficient because of the strong CO depletion that results 
from high densities.
Lower densities just outside the dense ring may lead to lower
CO depletion and a more efficient  DCO$^+$ formation, explaining
the DCO$^+$ peak there.

\paragraph{ Other observed species:} We also presented integrated column densities of HC$_3$N and $\it c$-C$_3$H$_2$, in Table \ref{table:dens} and 
Appendix \ref{apen:chem}. The modeled column densities of HC$_3$N and $\it 
c-$C$_3$H$_2$ are overpredicted. The high column density of HC$_3$N above three scale 
heights is due to its rapid formation via  \mbox{CN + C$_2$H$_2$$\rightarrow$ H + HC$_3$N} reaction, 
whereas $\it 
c-$C$_3$H$_2$ forms from the \mbox{CH + C$_2$H$_2$} reaction, 
photodissociation of CH$_2$CCH and dissociative recombination of 
C$_3$H$_5$$^+$.

\section{Summary}
Using NOEMA, we have observed the GG Tau A outer disk in several 
molecules. We report the first detection of H$_2$S in a protoplanetary 
disk.

We clearly detect HCO$^+$, H$^{13}$CO$^+$, DCO$^+$, and H$_2$S.  
HCO$^+$ emission is extended, H$^{13}$CO$^+$ and H$_2$S emissions peak 
inside the dense ring at $\sim$ 250 au, while DCO$^+$ 
emission arises from the outer disk beyond a radius of 300 au,
perhaps as a result of competition between CO depletion and
high temperatures.

Our three-phase chemical model fails to reproduce the observed 
column densities of S-bearing molecules, even with low S abundance and 
C/O = 0.7, suggesting that our understanding of S chemistry on dust 
grains is still incomplete.

Comparisons with other disks indicate that the detection of 
H$_2$S appears to be facilitated by the large disk mass, but that the
relative abundance ratios remain similar. This indicates that GG Tau A
could be a good test bed for chemistry in disks.
    
 \begin{acknowledgements}
 We thank the referee for useful comments that helped improve the quality of the manuscript.
This work is based on observations carried out with the IRAM NOEMA Interferometer. IRAM is supported by INSU/CNRS (France), MPG (Germany) and IGN (Spain). This work has made use of data from the European Space Agency (ESA) mission
{\it Gaia} (\url{https://www.cosmos.esa.int/gaia}), processed by the {\it Gaia}
Data Processing and Analysis Consortium (DPAC,
\url{https://www.cosmos.esa.int/web/gaia/dpac/consortium}). Funding for the DPAC
has been provided by national institutions, in particular the institutions
participating in the {\it Gaia} Multilateral Agreement.
Anne Dutrey and St\'ephane Guilloteau thank the French CNRS programs PNP, PNPS, and PCMI.
N. T. Phuong and P. N. Diep acknowledge financial support from NAFOSTED under grant number 103.99-2016.50, World Laboratory, Rencontres du Viet Nam, the Odon Vallet fellowships, Vietnam National Space Center, and Graduate University of Science and Technology. V. Wakelam's research is funded by an ERC Starting Grant (3DICE, grant agreement 336474).
\end{acknowledgements}

\bibliography{references}
\bibliographystyle{aa}


\begin{appendix} 
\section{Channel maps}
\label{apen:chan}
 \begin{figure}[htbp!]
   \centering
   \includegraphics[width=\hsize, trim=1cm 6cm 5.3cm 2cm]{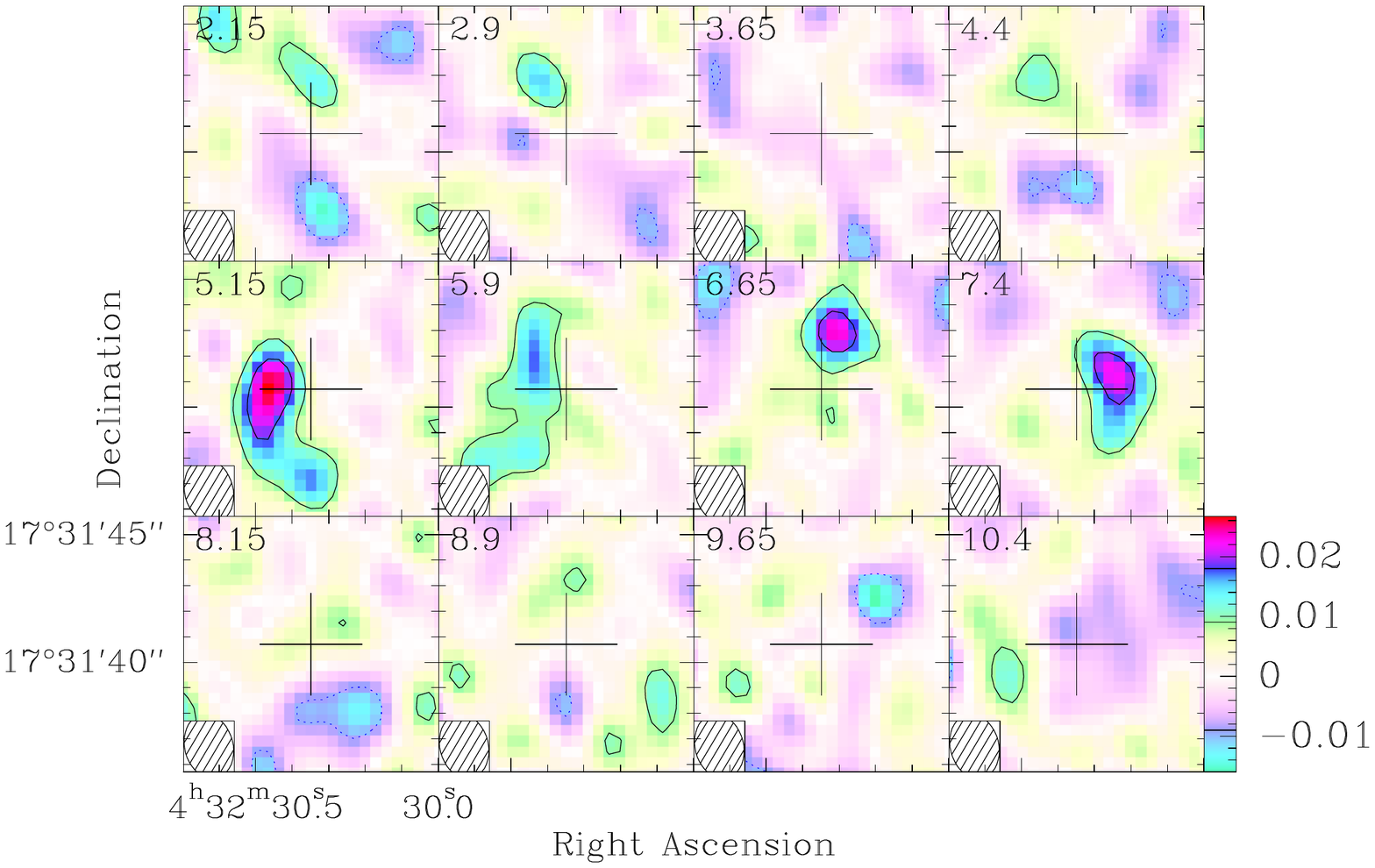}
      \caption{Channel maps H$_2$S 1(1,0) - 1(0,1) emission. The color scale is in the unit of Jy\,beam$^{-1}$. The contour spacing is 10\, mJy\,beam$^{-1}$, which corresponds to 2$\sigma$ or 0.08\,K. 
      The beam ($2.55"\times1.90"$, PA=14$^\circ$) is inserted in the lower corner of each channel map.}
         \label{h2s}
   \end{figure}
 %
 \begin{figure}[htbp!]
   \centering
   \includegraphics[width=\hsize, trim=1cm 6cm 5.3cm 2cm]{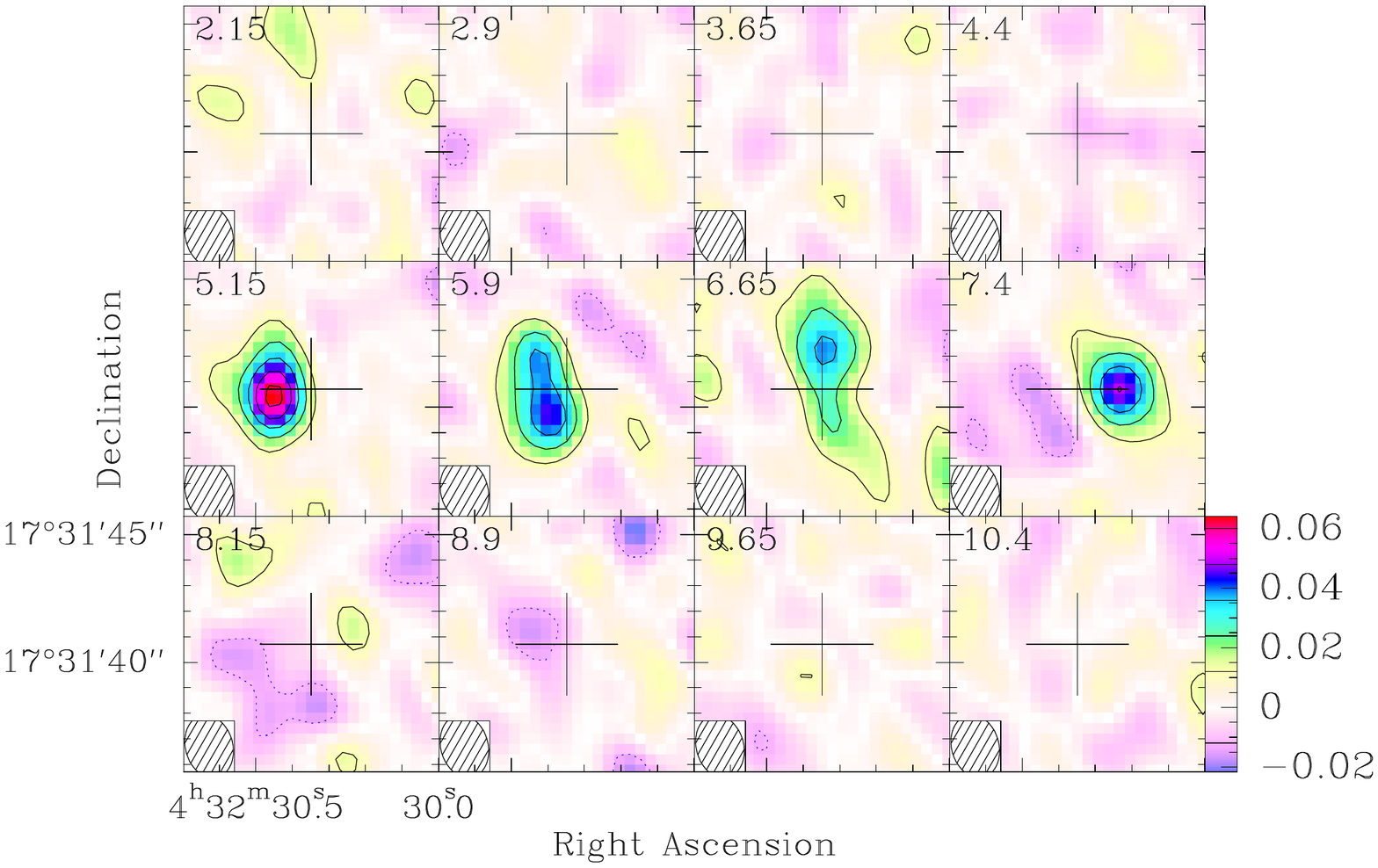}
      \caption{Channel maps H$^{13}$CO$^+$ (2-1) emission. The color scale is in the unit of Jy\,beam$^{-1}$. The contour spacing is 12\, mJy\,beam$^{-1}$, which corresponds to 2$\sigma$ or 0.11\,K. 
      The beam ($2.50"\times1.85"$, PA=15$^\circ$) is inserted in the lower corner of each channel map.}
         \label{h13cop}
   \end{figure}    
  %
 \begin{figure}[htbp!]
   \centering
   \includegraphics[width=\hsize,trim=1cm 6cm 5.3cm 2cm]{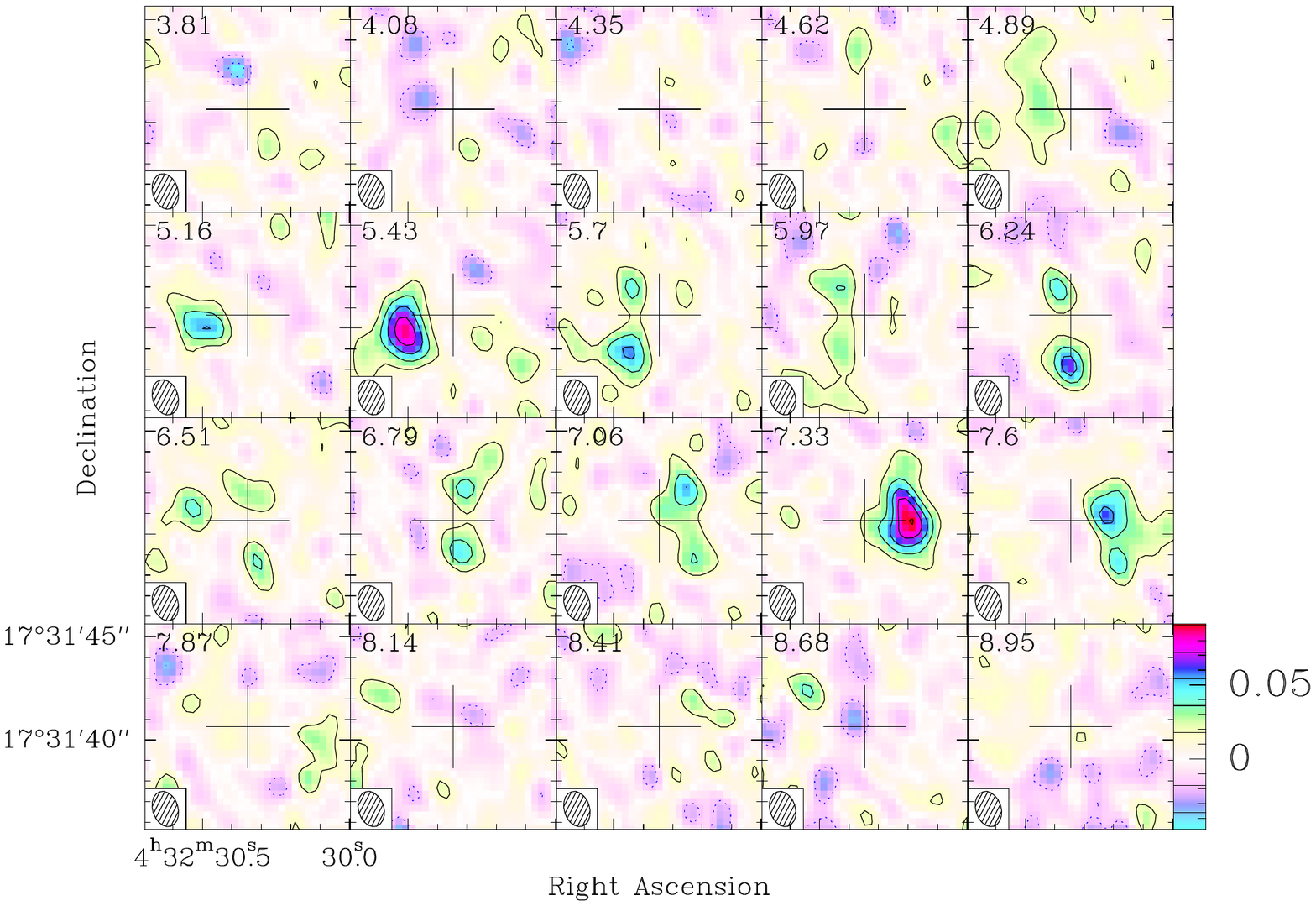}
      \caption{Channel maps DCO$^+$ (3-2) emission. The color scale is in the unit of Jy\,beam$^{-1}$. The contour spacing is 18\, mJy\,beam$^{-1}$, which corresponds to 2$\sigma$ or 0.22\,K. 
      The beam ($1.76"\times1.23"$, PA=17$^\circ$) is inserted in the lower corner of each channel map.}
         \label{dcop}
   \end{figure}
 %
 \begin{figure}[htbp!]
   \centering
   \includegraphics[width=\hsize,trim=1cm 6cm 5.3cm 2cm]{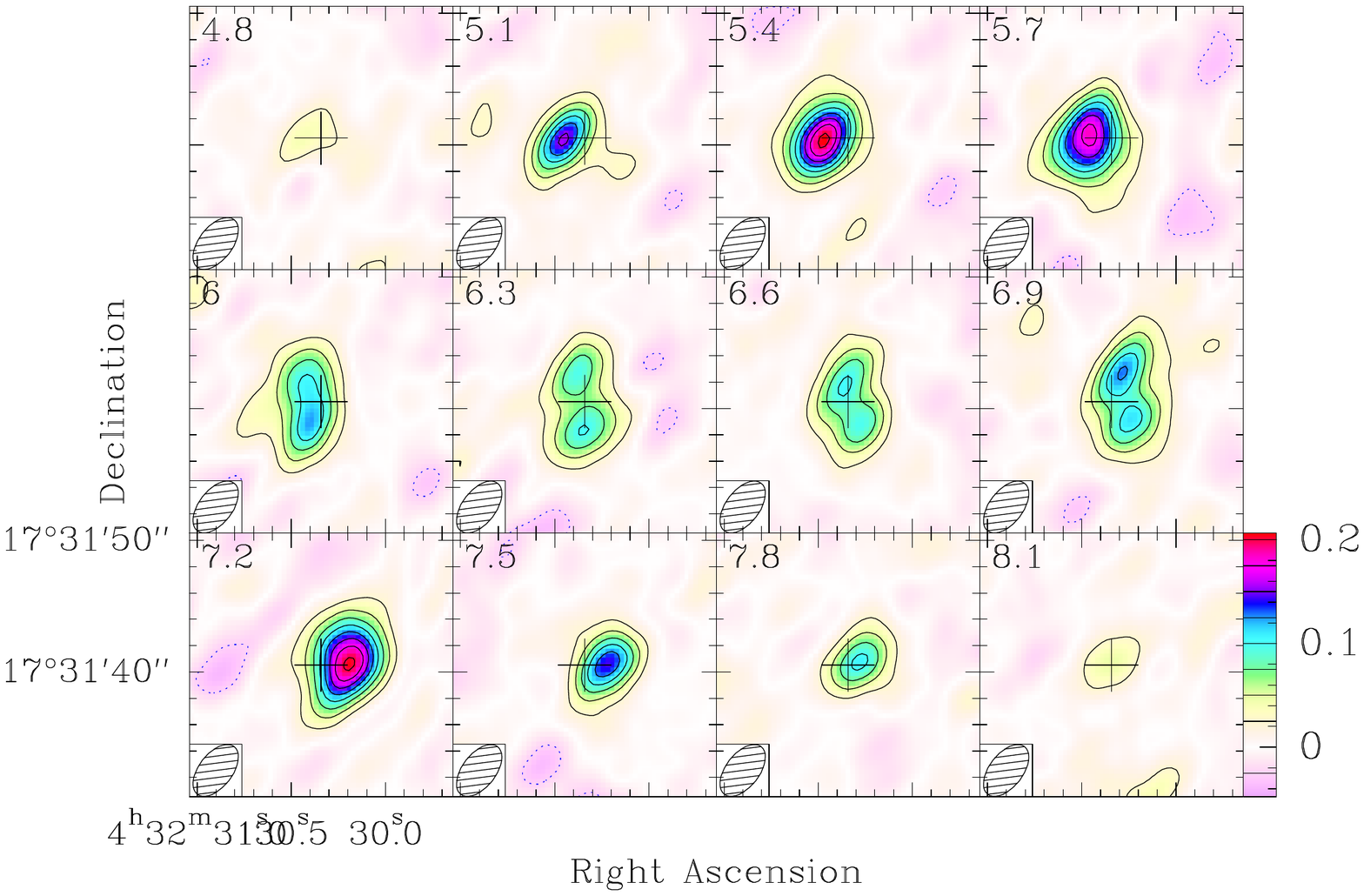}
      \caption{Channel maps HCO$^+$ (1-0) emission. The color scale is in the unit of Jy\,beam$^{-1}$. The contour spacing is 25\, mJy\,beam$^{-1}$, which corresponds to 2$\sigma$ or 0.33\,K. 
      The beam ($4.57"\times2.55"$, PA=$-38^\circ$) is inserted in the lower corner of each channel map.}
         \label{hcop}
   \end{figure}

\newpage
\section{Vertical integrated molecule column densities}
\label{apen:chem}
\begin{figure}[htbp!]
   \centering
  \includegraphics[width=2.8cm, angle=90]{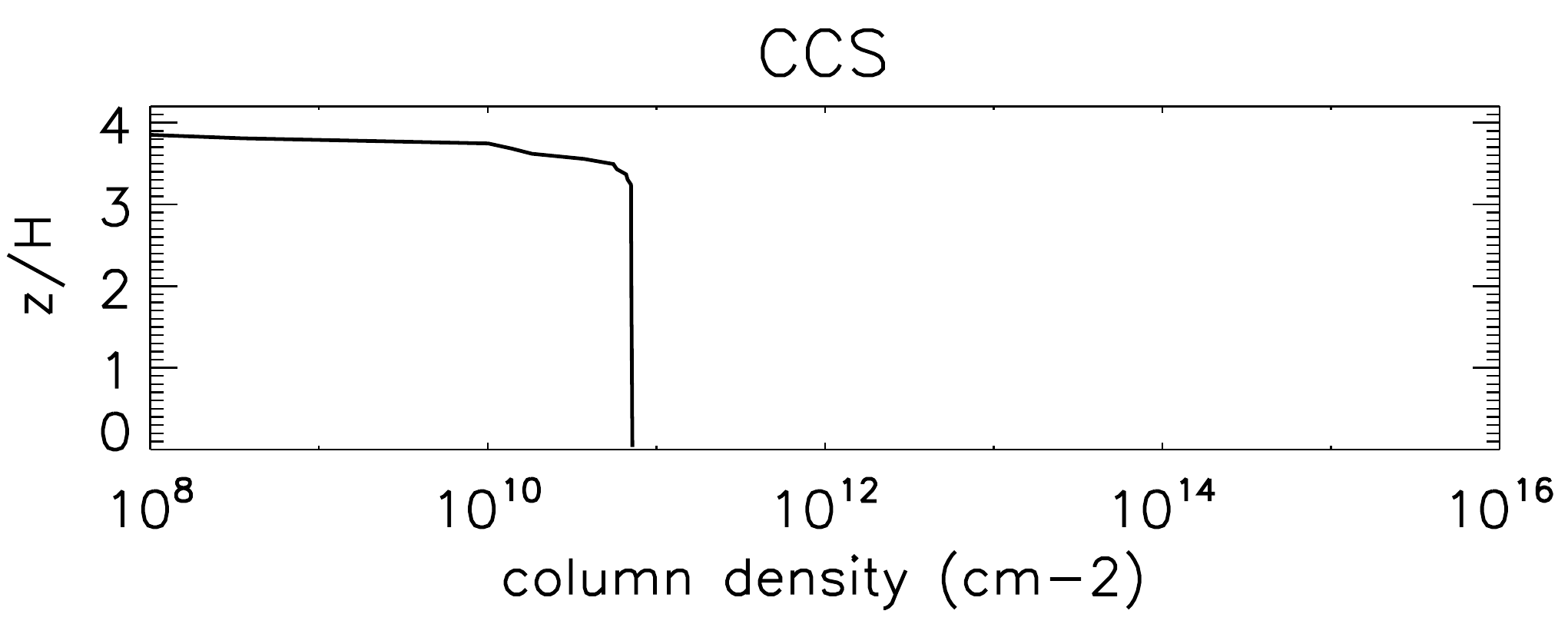}\\
  \includegraphics[width=2.8cm, angle=90]{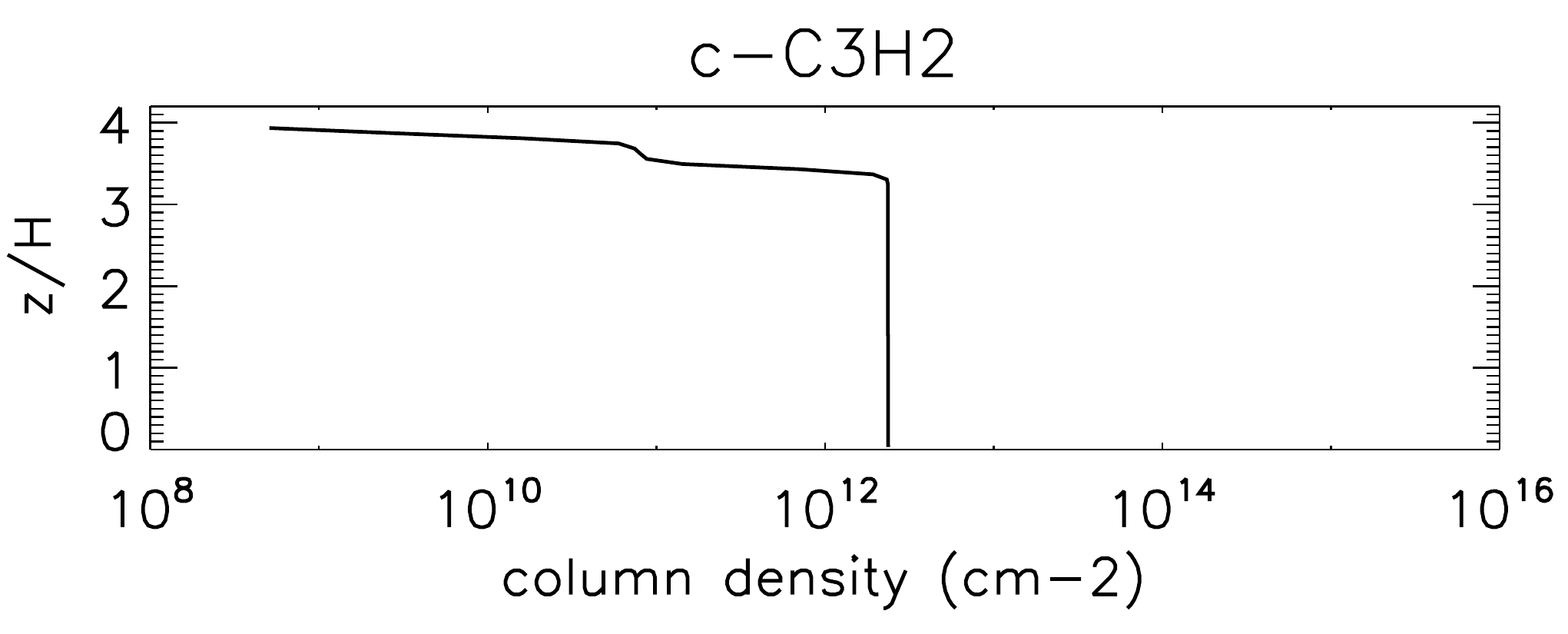}\\
  \includegraphics[width=2.8cm, angle=90]{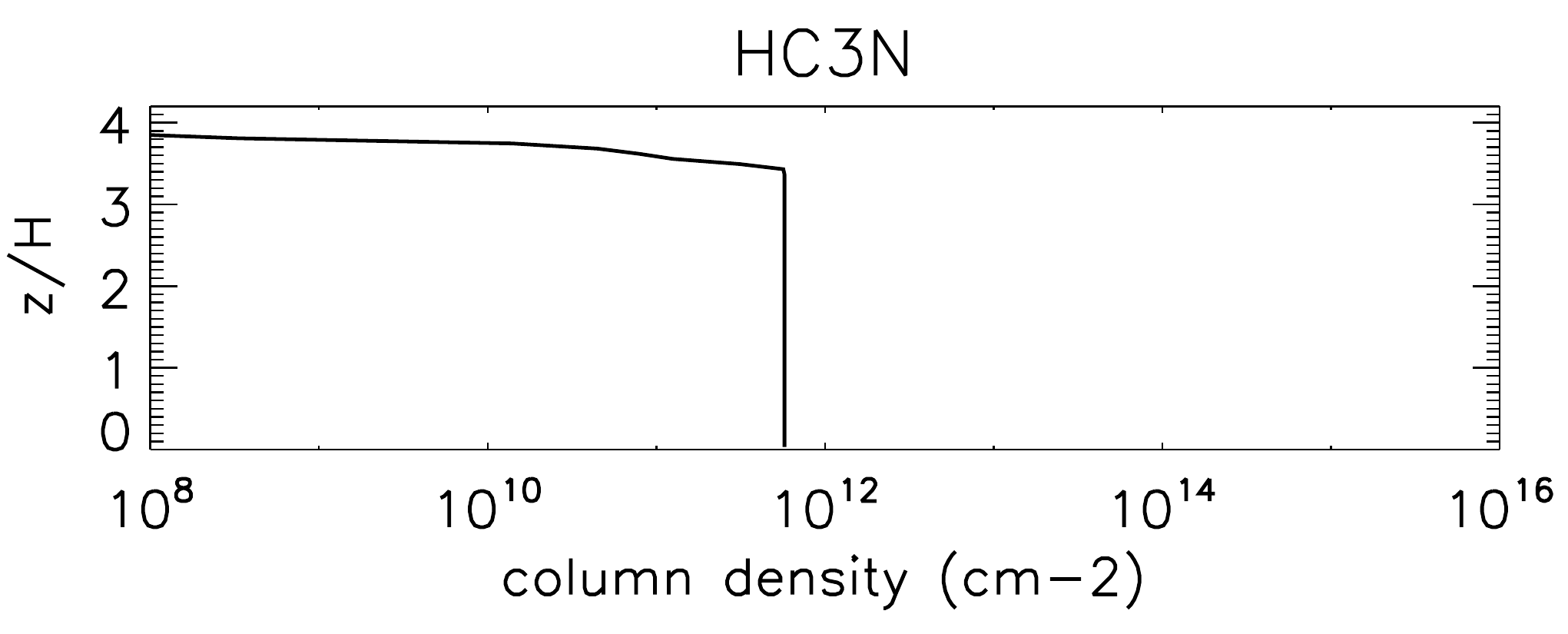}\\
  \includegraphics[width=2.8cm, angle=90]{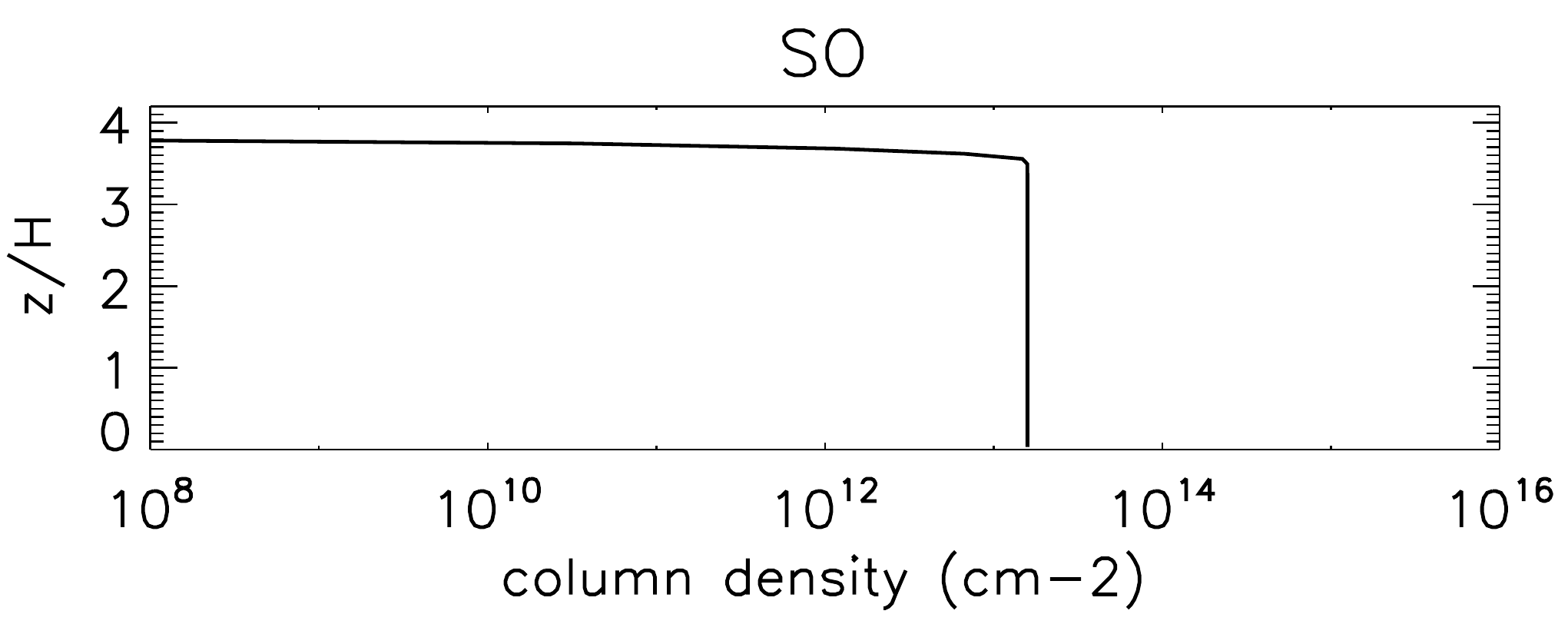}\\
  \includegraphics[width=2.8cm, angle=90]{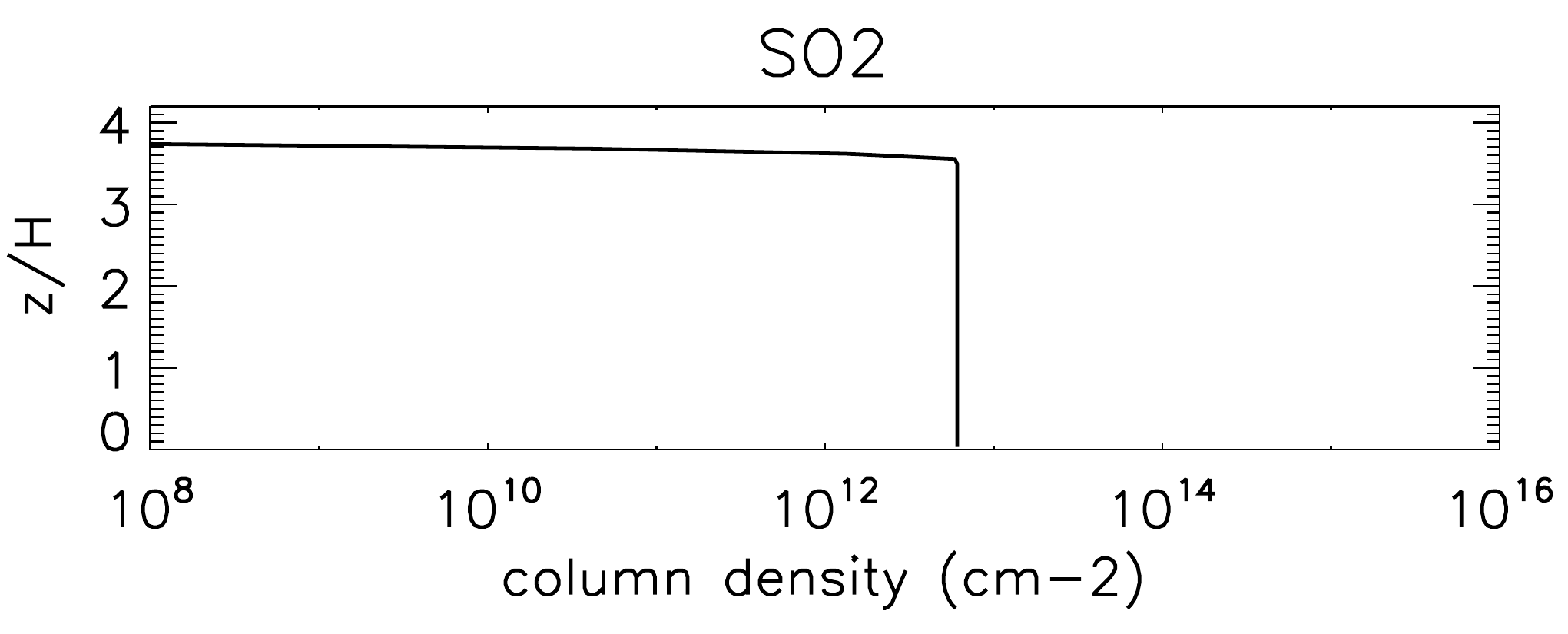}\\
   \caption{Best model of CCS, c-C$_3$H$_2$, HC$_3$N, SO, and SO$_2$ in the GG Tau A ring, derived from Nautilus, using our best knowledge of the GG Tau disk.}
              \label{Fig2}
    \end{figure}    
\end{appendix}

%
%

\end{document}